\documentclass[aps,pre,reprint,floatfix,nofootinbib]{revtex4-2}
\usepackage[utf8]{inputenc}
\usepackage{siunitx}
\usepackage{graphicx}
\usepackage{physics}
\usepackage{amsmath}
\usepackage{amssymb}
\usepackage{bbm}
\usepackage{bm}
\usepackage{dsfont}
\usepackage{multirow}
\usepackage[normalem]{ulem}
\usepackage[ruled,vlined,linesnumbered]{algorithm2e}
\usepackage[dvipsnames]{xcolor}
\usepackage[normalem]{ulem}

\newcommand{\dir}{\bm{\hat{\Omega}}}
\newcommand{\pos}{\bm{r}}

\newcommand{\Et}{\Sigma_t}
\newcommand{\Es}{\Sigma_s}

\newcommand{\Emaj}{\Sigma_\text{maj}}
\newcommand{\keff}{k_{\text{eff}}}
\newcommand{\dirto}[2]{\frac{#2 - #1}{\abs{#2 - #1}}}

\newcommand{\tripoli}{{\sc TRIPOLI-4}\textsuperscript{\textregistered}}
\newcommand{\apollo}{{\sc APOLLO3}\textsuperscript{\textregistered}}

\DeclareMathOperator{\atantwo}{atan2}

\begin{document}
    \title{Variance Reduction and Noise Source Sampling Techniques for Monte Carlo Simulations of Neutron Noise Induced by Mechanical Vibrations}
    
    \author{Hunter Belanger}
    \email{hunter.belanger@cea.fr}
    \author{Davide Mancusi}
    \email{davide.mancusi@cea.fr}
    \author{Amélie Rouchon}
    \email{amelie.rouchon@cea.fr}
    \author{Andrea Zoia}
    \email{andrea.zoia@cea.fr}
    
    \affiliation{Universit\'e Paris-Saclay, CEA, Service d'\'Etudes des R\'eacteurs et de Math\'ematiques Appliqu\'ees, 91191, Gif-sur-Yvette, France}
    
    \begin{abstract}
        Neutron noise in nuclear power reactors refers to the small fluctuations around the average neutron flux at steady state resulting from time-dependent perturbations inside the core. The neutron noise equations in the frequency domain can be solved using Monte Carlo simulation codes, which are capable of obtaining reference solutions involving almost no approximations, but are hindered by severe issues affecting the statistical convergence: the simultaneous presence of positive and negative particles, which is required by the nature of the complex noise equations, leads to catastrophically large variance in the tallies. In this work, we consider the important case of neutron noise problems induced by mechanical vibrations. First, we derive a new exact sampling strategy for the noise source. Then, building upon our previous findings in other contexts, we show that weight cancellation methods can be highly beneficial in dealing with the presence of negative weights, enabling extremely large gains in the figure of merit. We successfully demonstrate our results on a benchmark configuration consisting of a fuel assembly with a vibrating pin and we discuss possible pathways for further improvements.
    \end{abstract}
    
    \maketitle
    
    \section{Introduction}
    \label{sec:intro}
    
    Neutron noise in nuclear power reactors is defined as the small fluctuations occurring around the average neutron flux at steady state due to perturbations typically induced by fluid-structure interactions or moderator density fluctuations inside the core \cite{williams, thie, pazsit_demaziere}. These perturbations may affect different reactor components, at the scale of single or multiple fuel rods, entire assemblies, or the full core vessel, depending on the specific physical origin of the disturbances. Generally speaking, neutron noise is an unwanted phenomenon, which in some extreme cases might lead to a significant decrease in reactor power in order to ensure safe operation, or even to the shut-down of the reactor~\cite{seidel_pwr}. Notwithstanding, such disturbances carry an information content that can be usefully exploited to improve reactor diagnostic through the application of inverse problem techniques, for instance by locating anomalous control rod or fuel assembly vibrations, or monitoring moderator speed and void fraction~\cite{behriguer, fry, pazsit_analytis, kosaly, pazsit_void, pazsit_ringhals}.
    
    The neutron noise equations are established by considering the effect of temperature and density changes and/or material displacements on the operators occurring in the Boltzmann equation and in the precursor equations: the noise is considered as a small time-dependent perturbation with respect to the stationary flux, and the time-dependent operators are similarly decomposed into a stationary part and a residual time-dependent perturbation~\cite{pazsit_demaziere}. Then, assuming that the products of perturbed quantities can be neglected (the so-called `orthodox' linearization~\cite{pazsit_linearization}), a Fourier transform finally leads to a complex-valued equation for the neutron noise in the frequency domain~\cite{pazsit_demaziere}.
    
    A number of analytical and semi-analytical results concerning the solutions of the noise equations have been obtained using diffusion theory, simplified few-group transport equations, and Green's function methods~\cite{pazsit_strong_epsd, jonsson, yamamoto_modes, pazsit_modes, Zoia2021, Demaziere2022}: although these approaches shed light on the behaviour of the neutron noise field in a variety of physical configurations, their application to real-world systems is somewhat limited. Recently, a renewed interest in neutron noise stimulated by industrial problems related to core diagnostics has fostered the development of novel numerical codes capable of solving the noise equations in the time or frequency domain using state-of-the-art methods~\cite{rohde, demaziere_coresim, Yamamoto2013, Rouchon2017, olmojouan, vidal, chionis}. In this respect, a remarkable contribution has been provided by the EU H2020 CORTEX project (2017-2021)~\cite{demaziere_physor}, during which several deterministic and Monte Carlo solvers for the noise equation have been conceived~\cite{yi_mc2021, yamamoto_bwr, rouchon_mc2019} and validated against benchmark problems~\cite{vinai_mc2019, Vinai2021, VinaiSubmitted2022} and experimental data stemming from dedicated measurement campaigns in research reactors~\cite{akr2_crocus, colibri}.
    
    Since the number of available experimental datasets for the qualification of faster, albeit approximate, deterministic solvers for the noise equations is still rather small, one would like to rely on Monte Carlo simulation as a gold-standard tool to compute reference solutions involving almost no approximations, similarly to what is customarily done for the regular Boltzmann equation in reactor physics and radiation shielding problems~\cite{Vinai2021, VinaiSubmitted2022}. Monte Carlo sampling methods devoted to the noise equations have been devised and successfully tested in several applications, but numerical investigations have shown that their statistical convergence may be extremely poor in some cases: this is basically due to the fact that solving the noise equations requires sampling particles carrying complex statistical weights with positive and negative real and imaginary components~\cite{Yamamoto2013, Rouchon2017}. When positive and negative contributions are tallied in order to estimate the neutron noise, the variance might become overwhelmingly large and a huge number of sampled particles is mandatory to achieve a reasonable statistical uncertainty, which makes such simulations unwieldy even for relatively simple configurations~\cite{Vinai2021, VinaiSubmitted2022}.
    
    The possibility of using Monte Carlo simulations as a routine tool for industrial calculations in the field of neutron noise therefore crucially depends on the availability of ad hoc variance reduction and population control techniques capable of handling the simultaneous presence of positive and negative particles. Recent insights have shown that weight cancellation methods might be the key to successfully dealing with such issues~\cite{Yamamoto2013,Belanger2021, BelangerMC2021}. Preliminary investigations involving cross section oscillation problems in a fuel assembly suggest that the introduction of weight cancellation actually leads to astounding improvements in the figure of merit of the noise simulations by Monte Carlo codes, at the expense of extensive restructuring of the sampling routines~\cite{BelangerPHYSOR2022}. The benefits induced by the vastly reduced statistical uncertainty establish Monte Carlo simulations as a practical tool to obtain reference solutions and outweighs the efforts involved in code rewriting. In this work, we extend and considerably generalize our previous findings to the case of neutron noise induced by mechanical vibrations, which poses distinct challenges in both sampling the noise source particles and handling the variance of the resulting noise field.

    This work is organized as follows. In Sec.~\ref{sec:noise_eq} we recall the derivation of the neutron noise equations in the frequency domain, and we show that they can be interpreted as a complex-valued fixed-source transport problem, starting from a complex-valued source. Subsequently, in Sec.~\ref{sec:sampling_source} we review existing sampling methods for the noise source, and introduce a new exact sampling strategy for the case of mechanical vibrations. After briefly recalling in Sec.~\ref{sec:mc_nn} the Monte Carlo methods for solving the transport of noise particles, in Sec.~\ref{sec:cancellation} we discuss the application of weight cancellation to reduce the variance of noise problems. The effectiveness of these methods is illustrated in Sec.~\ref{sec:simulation_results} on a benchmark configuration consisting of a fuel assembly where a pin is subject to periodic mechanical vibrations. Conclusions are finally drawn in Sec.~\ref{sec:conclusions}.
    
    \section{The Neutron Noise Equations}
    \label{sec:noise_eq}
    We start by considering a multiplying system which is initially critical. To allow for possible model bias due to technological uncertainties and/or nuclear data, we assume that the fission production is normalized by the fundamental eigenvalue $\keff$, which enforces the stationary state. Now, if a perturbation is introduced into the system, so that the macroscopic quantities such as cross sections, yields, and scattering laws become time-dependent, then the resulting neutron flux $\varphi(\pos,\dir,E,t)$ will also be time-dependent, and will be therefore described by the time-dependent Boltzmann equation, coupled to the evolution equations for the delayed neutron precursors. After injecting the precursor equations into the Boltzmann equation, we concisely denote the time-dependent transport equation as $\mathcal{B}_k(t)\varphi(\pos,\dir,E,t) = 0$, where
    \begin{widetext}
    \begin{multline}
        \mathcal{B}_k(t) =
        \frac{1}{v}\frac{\partial}{\partial t}+
        \dir\cdot\grad + \Et(\pos,E,t) -
        \iint \nu_s(\pos,E',t)
              \Sigma_s(\pos,E',t)
              f_s(E'\to E,\dir'\to\dir,t)
              \dd E'
              \dd\dir' - \\
        \frac{1}{\keff}
        \iint \nu_{f,p}(\pos,E',t)
              \Sigma_f(\pos,E',t)
              f_{f,p}(E'\to E,\dir'\to\dir,t)
              \dd E'
              \dd\dir' - \\
        \frac{1}{\keff}
        \sum_j
        \iiint \lambda_j e^{-\lambda_j(t-t')}
               \nu_{f,d}^j(\pos,E',t')
              \Sigma_f(\pos,E',t')
              f_{f,d}^j(E'\to E,\dir'\to\dir,t')
              \dd E'
              \dd\dir'
              \dd t'
        \text.
        \label{eq:time_boltzmann_pert}
    \end{multline}
    \end{widetext}
    Due to the effect of the perturbation, it is assumed that all of the time-dependent terms in Eq.~\eqref{eq:time_boltzmann_pert} can be be decomposed into a stationary part and an additional small, periodic, time-dependent contribution\footnote{In this work, we do not consider the possibility of perturbations in the precursor decay constants.}. For example, the total macroscopic cross section is decomposed as
    \begin{equation}
        \Et(\pos,E,t) = \Et(\pos,E) + \delta\Et(\pos,E,t)
        \text.
        \label{eq:Et_perturb}
    \end{equation}
    This can similarly be done for reaction yields, scattering laws and fission spectra. As it would become slightly cumbersome to write each integral explicitly, we use the following shorthand for the decomposition of a generic term $\alpha$:
    \begin{multline}
        \nu_\alpha(\pos,E,t)
        \Sigma_\alpha(\pos,E,t)
        f_\alpha(E\to E',\dir\to\dir',t) = \\
        \nu_\alpha(\pos,E)
        \Sigma_\alpha(\pos,E)
        f_\alpha(E\to E',\dir\to\dir') \\
        +\delta[\nu_\alpha\Sigma_\alpha f_\alpha](\pos,E\to E',\dir\to\dir',t)
        \text,
        \label{eq:full_perturbation_time}
    \end{multline}
    where all of the time dependence has been placed in the single $\delta[\nu_\alpha\Sigma_\alpha f_\alpha]$ term. A thorough discussion on this decomposition will be provided in Sec.~\ref{sec:sampling_source}. In this formalism, we may break the operator $\mathcal{B}_k(t)$ into two different components
    \begin{equation}
        \mathcal{B}_k(t) = \mathcal{B}(t) + \delta\mathcal{B}(t)
        \text,
    \end{equation}
    with the operator
    \begin{widetext}
    \begin{multline}
        \mathcal{B}(t) = 
        \frac{1}{v}\frac{\partial}{\partial t}+
        \dir\cdot\grad + \Et(\pos,E) -
        \iint \nu_s(\pos,E')
              \Sigma_s(\pos,E')
              f_s(E'\to E,\dir'\to\dir)
              \dd E'
              \dd\dir' - \\
        \frac{1}{\keff}
        \iint \nu_{f,p}(\pos,E')
              \Sigma_f(\pos,E')
              f_{f,p}(E'\to E,\dir'\to\dir)
              \dd E'
              \dd\dir' - \\
        \frac{1}{\keff}
        \sum_j
        \iiint \lambda_j e^{-\lambda_j(t-t')}
               \nu_{f,d}^j(\pos,E')
              \Sigma_f(\pos,E')
              f_{f,d}^j(E'\to E,\dir'\to\dir)
              \dd E'
              \dd\dir'
              \dd t'
        \text,
    \end{multline}
    and the perturbation operator
    \begin{multline}
       \delta\mathcal{B}(t) = 
       \delta\Et(\pos,E,t) -
        \iint 
            \delta[\nu_s\Sigma_s f_s](\pos,E'\to E,\dir'\to\dir,t)
            \dd E'
            \dd\dir' - \\
        \frac{1}{\keff}
        \iint
            \delta[\nu_{f,p}\Sigma_{f,p} f_{f,p}](\pos,E'\to E,\dir'\to\dir,t)
            \dd E'
            \dd\dir' - \\
        \frac{1}{\keff}
        \sum_j
        \iiint
            \lambda_j e^{-\lambda_j(t-t')}
            \delta[\nu_{f,d}^j\Sigma_{f,d} f_{f,d}^j](\pos,E'\to E,\dir'\to\dir,t')
            \dd E'
            \dd\dir'
            \dd t'
        \text.
    \end{multline}
    \end{widetext}
    Consequently, we postulate that the time-dependent flux which solves Eq.~\eqref{eq:time_boltzmann_pert} can be similarly decomposed as
    \begin{equation}
        \varphi(\pos,\dir,E,t) = \varphi_c(\pos,\dir,E) + \delta\varphi(\pos,\dir,E,t)
        \text,
    \end{equation}
    where $\varphi_c$ is the fundamental eigenmode of the Boltzmann equation associated to the fundamental eigenvalue $\keff$, before the time-dependent disturbance was introduced, and the flux perturbation $\delta\varphi$ is the neutron noise. From these definitions, we see that
    \begin{align}
        \mathcal{B}_k&(t)\varphi(\pos,\dir,E,t) = 0 \nonumber\\
        & =\bigg[\mathcal{B}(t) + \delta\mathcal{B}(t)\bigg]
        \bigg[\varphi_c(\pos,\dir,E) + \delta\varphi(\pos,\dir,E,t)\bigg]  \nonumber\\
        & = \mathcal{B}(t)\delta\varphi(\pos,\dir,E,t) +
        \delta\mathcal{B}(t)\varphi_c(\pos,\dir,E) \nonumber\\
        & + \mathcal{B}(t)\varphi_c(\pos,\dir,E) + 
        \delta\mathcal{B}(t)\delta\varphi(\pos,\dir,E,t)
        \text.
    \end{align}
    The term $\mathcal{B}(t)\varphi_c(\pos,\dir,E)$ vanishes, since $\varphi_c$ is the critical (stationary) flux. Next, we apply the `orthodox' linearization, where we assume that for small perturbations the terms involving products of perturbed quantities can be neglected, namely, $\delta\mathcal{B}\delta\varphi\approx 0$. The validity of this assumption has been discussed by several authors~\cite{pazsit_linearization, pazsit_epsd,pazsit_strong_epsd}, and has been recently revisited for the case of mechanical vibrations \cite{Zoia2021,Demaziere2022}. Upon rearranging the two remaining terms, we obtain a Boltzmann-like transport equation for the neutron noise which resembles a fixed-source problem:
    \begin{equation}
        \mathcal{B}(t)\delta\varphi(\pos,\dir,E,t) =
        -\delta\mathcal{B}(t)\varphi_c(\pos,\dir,E)
        \text,
        \label{eq:noise_time}
    \end{equation}
    the term $-\delta\mathcal{B}\varphi_c$ representing the `noise source'.
    
    Since we are typically interested in the periodic solution to the noise equation, we apply then a Fourier transform
    \begin{equation}
    g(\omega) = {\mathcal F} \left[ g(t)\right](\omega)  = \int_{-\infty}^{+\infty} e^{-i\omega t} g(t) dt
    \end{equation}
    to Eq.~\eqref{eq:noise_time}, $\omega$ denoting the angular frequency, which yields the linearized noise equation in the frequency domain:
    \begin{equation}
        \mathcal{B}(\omega)\delta\varphi(\pos,\dir,E,\omega) =
        -\delta\mathcal{B}(\omega)\varphi_c(\pos,\dir,E)
        \text.
        \label{eq:noise_freq}
    \end{equation}
    The Fourier-transformed operators appearing in Eq.~\eqref{eq:noise_freq} are complex-valued and read
    \begin{widetext}
    \begin{multline}
        \mathcal{B}(\omega) = 
        i\frac{\omega}{v} +
        \dir\cdot\grad + \Et(\pos,E) -
        \iint \nu_s(\pos,E')
              \Sigma_s(\pos,E')
              f_s(E'\to E,\dir'\to\dir)
              \dd E'
              \dd\dir' - \\
        \frac{1}{\keff}
        \iint \nu_{f,p}(\pos,E')
              \Sigma_f(\pos,E')
              f_{f,p}(E'\to E,\dir'\to\dir)
              \dd E'
              \dd\dir' - \\
        \frac{1}{\keff}
        \sum_j
        \frac{\lambda_j}{\lambda_j + i\omega} \iint 
              \nu_{f,d}^j(\pos,E')
              \Sigma_f(\pos,E')
              f_{f,d}^j(E'\to E,\dir'\to\dir)
              \dd E'
              \dd\dir'
        \text,
        \label{eq:noise_boltzmann}
    \end{multline}
    and
    \begin{multline}
       \delta\mathcal{B}(\omega) = 
       \delta\Et(\pos,E,\omega) -
        \iint 
            \delta[\nu_s\Sigma_s f_s](\pos,E'\to E,\dir'\to\dir,\omega)
            \dd E'
            \dd\dir' - \\
        \frac{1}{\keff}
        \iint
            \delta[\nu_{f,p}\Sigma_{f,p} f_{f,p}](\pos,E'\to E,\dir'\to\dir,\omega)
            \dd E'
            \dd\dir' - \\
        \frac{1}{\keff}
        \sum_j
         \frac{\lambda_j}{\lambda_j + i\omega}
        \iint
            \delta[\nu_{f,d}^j\Sigma_{f,d} f_{f,d}^j](\pos,E'\to E,\dir'\to\dir,\omega)
            \dd E'
            \dd\dir'
        \text.
        \label{eq:noise_source_fourier}
    \end{multline}
    \end{widetext}
    For symmetry, it is sometimes convenient to write the $\delta\Et(\pos,E,\omega)$ term in Eq.~\eqref{eq:noise_source_fourier} in the same form as the other terms:
    \begin{multline}
        \delta\Et(\pos,E,\omega) = \\
        \iint
        \delta\Et(\pos,E',\omega)
        \delta(E'-E)
        \delta(\dir'-\dir)
        \dd E'
        \dd\dir'
        \text,
        \label{eq:copy_source_definition}
    \end{multline}
    which shows that $\delta\Et(\pos,E,\omega)$ is associated to a unit yield and a `delta-copy' spectrum $\delta(E'-E)\delta(\dir'-\dir)$. The terms $\delta\Et(\pos,E,\omega)$ and $\delta[\nu_\alpha\Sigma_\alpha f_\alpha](\pos,E'\to E,\dir'\to\dir,\omega)$ appearing in the noise source involve the Fourier-transformed perturbation of the transport operators, and as such their precise functional form is determined by the specific type of perturbation of the noise model (e.g., mechanical vibrations or oscillations). Regardless of the nature of the solver (deterministic or Monte Carlo), these quantities must thus be prepared in advance, in the source code, before solving the noise equation~\eqref{eq:noise_freq}. In the next section we will discuss the exact form of the noise source for two different types of perturbations, and the Monte Carlo strategies that we have implemented to sample them.
   
    \section{Sampling the Neutron Noise Source}
    \label{sec:sampling_source}
    Sampling the neutron noise source requires the knowledge of the Fourier-transformed perturbation operator $-\delta\mathcal{B}(\omega)$ and the fundamental eigenmode $\varphi_c(\pos,\dir,E)$. The fundamental eigenmode can be obtained by running the well-known power iteration algorithm, which leads to a collection of neutrons whose flux is precisely $\varphi_c(\pos,\dir,E)$, after a sufficiently large numbers of inactive cycles. When convergence has been attained, $-\delta\mathcal{B}(\omega) \varphi_c(\pos,\dir,E)$ can be estimated by running one or more additional power iteration cycle, where the noise source particles are sampled from the histories of the neutrons, and are placed in a dedicated bank \cite{rouchon_mc2019}. This sampling strategy is inspired from the algorithm used to implement Monte Carlo kinetic simulations~\cite{Faucher2018}. The explicit form of $-\delta\mathcal{B}(\omega)$ depends as mentioned on the type of noise source to be examined.
    
    \subsection{Cross Section Oscillations}
    \label{sec:oscillations}
    
    First, we shall consider the simple case of cross section oscillations, which, although somewhat artificial, is nonetheless important in that it will play a key role in addressing the more involved case of mechanical vibrations, covered in Sec.~\ref{sec:mech_vibs}. Cross section oscillations are relatively straightforward, since they do not involve any spatial movement of the materials. Suppose there is a spatial region $\mathcal{P}$ where the macroscopic cross sections exhibit a periodic time dependence. We write the cross sections as \cite{Vinai2021}
    \begin{equation}
        \Sigma_\alpha(\pos,E,t) = \Sigma_\alpha(\pos,E) \left[1 + \varepsilon_\alpha\sin(\omega_0 t)\mathbbm{1}_\mathcal{P}(\pos)\right]
        \text,
    \end{equation}
    where $\varepsilon_\alpha$ is the amplitude of the perturbation, and
    \begin{equation}
       \mathbbm{1}_\mathcal{P}(\pos)  = 
       \begin{cases}
           1 & \text{for }\pos\in\mathcal{P} \\
           0 & \text{for }\pos\not\in\mathcal{P}
       \end{cases}
       \text.
    \end{equation}
    Upon taking the Fourier transform, the cross section perturbation in the frequency domain is then
    \begin{multline}
        \delta\Sigma_\alpha(\pos,E,\omega) =\\ 
        -i\pi
        \varepsilon_\alpha
        \Sigma_\alpha(\pos,E)
        \mathbbm{1}_\mathcal{P}(\pos)
        \left[\delta(\omega-\omega_0) + \delta(\omega + \omega_0)\right] \\
        = \Sigma_\alpha(\pos,E)\mathbbm{1}_\mathcal{P}(\pos) h_\alpha(\omega,\omega_0,\varepsilon_\alpha)
        \text.
    \end{multline}
    We assume that no other material properties display any time dependence. Such a problem is admittedly academic, but provides an excellent benchmark for comparison of codes and verification of algorithms. The noise source terms in Eq.~\eqref{eq:noise_source_fourier} can be thus written as
    \begin{multline}
        \iint
        \nu_\alpha(\pos,E')
        \delta\Sigma_\alpha(\pos,E',\omega)
        \times \\
        f_\alpha(\pos,E'\to E,\dir'\to\dir)
        \varphi_c(\pos,\dir',E')
        \dd E'
        \dd\dir'
        \text.
        \label{eq:oscil_source_term}
    \end{multline}
    By multiplying and dividing by $\Sigma_\alpha(\pos,E')$ in the integrand, and rearranging, we then have
    \begin{multline}
        \iint
        \underbrace{\frac{\delta\Sigma_\alpha(\pos,E',\omega)}{\Sigma_\alpha(\pos,E')}}_{\text{complex importance factor}} \times \\
        \underbrace{\nu_\alpha(\pos,E')
        \Sigma_\alpha(\pos,E')
        f_\alpha(\pos,E'\to E,\dir'\to\dir)}_{\text{standard production rate}}
        \times \\
        \varphi_c(\pos,\dir',E')
        \dd E'
        \dd\dir'
        \text.
        \label{eq:oscil_imoprt_source}
    \end{multline}
    Eq.~\eqref{eq:oscil_imoprt_source} can be handled by Monte Carlo codes with importance sampling, where we use the standard sampling methods for the production rate, and multiply the weight of the sampled noise particles corresponding to reaction $\alpha$ by the complex factor $\delta\Sigma_\alpha(\pos,E,\omega)/\Sigma_\alpha(\pos,E) = h_\alpha(\omega,\omega_0,\varepsilon_\alpha) \mathbbm{1}_\mathcal{P}(\pos)$. To sample the noise source, action only needs to be taken when a neutron has a collision within the region $\mathcal{P}$ exhibiting the cross section oscillations. At this point, we need to separately sample the three components of the noise source, that is, the copy term $\delta \Sigma_t \varphi_c$, the fission term, and the scattering term. The noise source contribution from the term $\delta \Sigma_t \varphi_c$ is sampled by creating an exact copy of the particle and multiplying its weight by $h_t(\omega,\omega_0,\varepsilon_t)$. For fission and scattering, a nuclide is sampled from the material at $\pos$, in the standard manner, and we apply forced fission and implicit absorption. If the nuclide is fissile, then a random variable $\xi\sim\mathcal{U}(0,1)$ is used to calculate the number of fission noise particles to produce: 
    \begin{equation}
       n = \left\lfloor \frac{\nu_f\left(E\right)\sigma_f\left(E\right)}{\sigma_t\left(E\right)\keff} + \xi \right\rfloor
       \text,
    \end{equation}
    where $\sigma_t(E)$ and $\sigma_s(E)$ are respectively the microscopic total and scattering cross sections for the selected nuclide. Each fission noise particle will inherit the weight of its parent, and have its energy and direction sampled normally. The weights of these particles are multiplied by $h_f(\omega,\omega_0,\varepsilon_f)$; furthermore, any delayed neutrons will also have their weight multiplied by the complex yield $\lambda_j/(\lambda_j+i\omega)$. Finally, noise particles from scattering must be sampled; first we sample a scattering channel, and then we sample the energy and direction of the scattered noise particle. In addition to multiplying the noise particle's weight by $h_s(\omega,\omega_0,\varepsilon_s)$, it is also necessary to multiply by the probability of scatter, $\sigma_s(E)/\sigma_t(E)$, where $\sigma_s(E)$ is the microscopic scattering cross sections for the selected nuclide. 
    
    \subsection{Mechanical Vibrations}\label{sec:mech_vibs}
    Mechanical vibrations are inherently different from the previous case of cross section oscillations, due to the fact that they involve material displacements as a function of time. Assuming that the displacement occurs along one spatial dimension, we will have material $M_L$ at position $x < x_0$ and material $M_R$ at position $x > x_0$ before the perturbation is applied. In stationary conditions, the material as a function of position is thus described by the function
    \begin{equation}
        M(x) = \begin{cases}
            M_L & \text{for }x < x_0 \\
            M_R & \text{for }x > x_0
        \end{cases}
        \text.
    \end{equation}
    Now, if a perturbation is introduced into this system in the form of a sinusoidal vibration of the interface between the two materials, with angular frequency $\omega_0$ and amplitude $\varepsilon > 0$, then the material found at any given position at a given position in the perturbed region is also a function of time. For positions where $\abs{x - x_0} < \varepsilon$, the function which determines the material at position $x$ and time $t$ is written as
    \begin{multline}
        M(x,t) = M(x) + \\
        \Delta M
        \left[H(x-x_0) - H(x-x_0-\varepsilon\sin(\omega_0 t))\right]
        \text,
        \label{eq:material_time}
    \end{multline}
    where $\Delta M = M_L - M_R$, and $H(x)$ is the Heaviside function \cite{Zoia2021}. This function always returns either $M_L$ or $M_R$. Based on the material perturbation $\delta M =  M(x,t) - M(x)$ across the interface, the corresponding perturbation of the operator kernels $\delta[\nu_\alpha\Sigma_\alpha f_\alpha]$ must then be
    \begin{multline}
        \delta[\nu_\alpha\Sigma_\alpha f_\alpha](x,E'\to E,\dir'\to\dir,t) = \Delta[\nu_\alpha\Sigma_\alpha f_\alpha] \times\\
        \left[H(x-x_0) - H(x-x_0-\varepsilon\sin(\omega_0 t))\right]
        \text,
        \label{eq:interface_time}
    \end{multline}
    where
    \begin{multline}
        \Delta[\nu_\alpha\Sigma_\alpha f_\alpha] = \\
        \nu_\alpha^L(E)\Sigma_\alpha^L(E)f_\alpha^L(E'\to E,\dir'\to\dir) -\\
        \nu_\alpha^R(E)\Sigma_\alpha^R(E)f_\alpha^R(E'\to E,\dir'\to\dir)
        \text.
    \end{multline}
    The Fourier transform of Eq.~\eqref{eq:interface_time} has been shown to be \cite{Zoia2021}
    \begin{multline}
        \delta[\nu_\alpha\Sigma_\alpha f_\alpha](x,E'\to E,\dir'\to\dir,\omega) = \\
        \Delta[\nu_\alpha\Sigma_\alpha f_\alpha]
        \bigg\{c_0(x,x_0)\delta(\omega) + \\
        \sum_{n=1}^\infty c_n(x,x_0)\left[\delta(\omega - n\omega_0) + \delta(\omega + n\omega_0)
        (-1)^n
        \right]
        \bigg\}
        \text,
        \label{eq:interface_ft}
    \end{multline}
    with
    \begin{equation}
       c_n(x,x_0) = \frac{2}{n}\sin(n\arcsin(\frac{x - x_0}{\varepsilon})) e^{-in\pi/2}
       \label{eq:c_n}
    \end{equation}
    for $n \ge 1$, and
    \begin{equation}
        c_0(x,x_0) =
        \begin{cases}
            \pi - 2\arcsin(\frac{x - x_0}{\varepsilon}) & \text{for } x \ge x_0 \\ \\
            -\pi - 2\arcsin(\frac{x - x_0}{\varepsilon}) & \text{for } x < x_0
        \end{cases}
        \text.
    \end{equation}
    Immediately apparent from Eq.~\eqref{eq:interface_ft} is that, while the noise source for cross section oscillations is monochromatic, the noise source for mechanical vibrations contains an infinite number of discrete harmonics at $n \omega_0$, all multiple of the perturbation frequency $\omega_0$. Additionally, there is a non-trivial spatial dependence indicated by Eq.~\eqref{eq:c_n}. All even harmonics are purely real and have even spatial parity about $x_0$, whereas all odd harmonics are purely imaginary and have odd parity. We will now consider possible methods to sample the noise source for mechanical vibrations in a Monte Carlo code.
    
    \subsubsection{Approximate Noise Source Sampling}
    \label{sec:approximated_source}
    
    From Eq.~\eqref{eq:interface_time}, we see that the noise source is caused by a difference of the product of the cross section, reaction yield, and scattering law. In the initial approach implemented in the noise solver of the \tripoli{} Monte Carlo code, developed at CEA \cite{T4}, it was assumed that
    \begin{equation}
        \Delta[\nu_\alpha\Sigma_\alpha f_\alpha] \approx
        \nu_\alpha f_\alpha \Delta\Sigma_\alpha(x,x_0,\omega)
        \text,
        \label{eq:t4_vib_source}
    \end{equation}
    where only the difference in the macroscopic cross sections was considered. It must be noted that, while $\Delta\Sigma_\alpha$ requires evaluating the cross section on both sides of the interface, $\nu_\alpha$ and $f_\alpha$ were both evaluated only for the material at the location where the collision occurred. This was partially done for simplicity, as evaluating $f_\alpha$ on both sides of the interface to obtain $\Delta[\nu_\alpha\Sigma_\alpha f_\alpha]$ could pose particular challenges\footnote{While the data representations used in Monte Carlo codes allow for easy sampling from the distribution $f_\alpha$, they often do not facilitate the evaluation of the distribution $f_\alpha$ corresponding to a given argument.}.
    
    \begin{figure}
        \centering
        \includegraphics[width=\columnwidth]{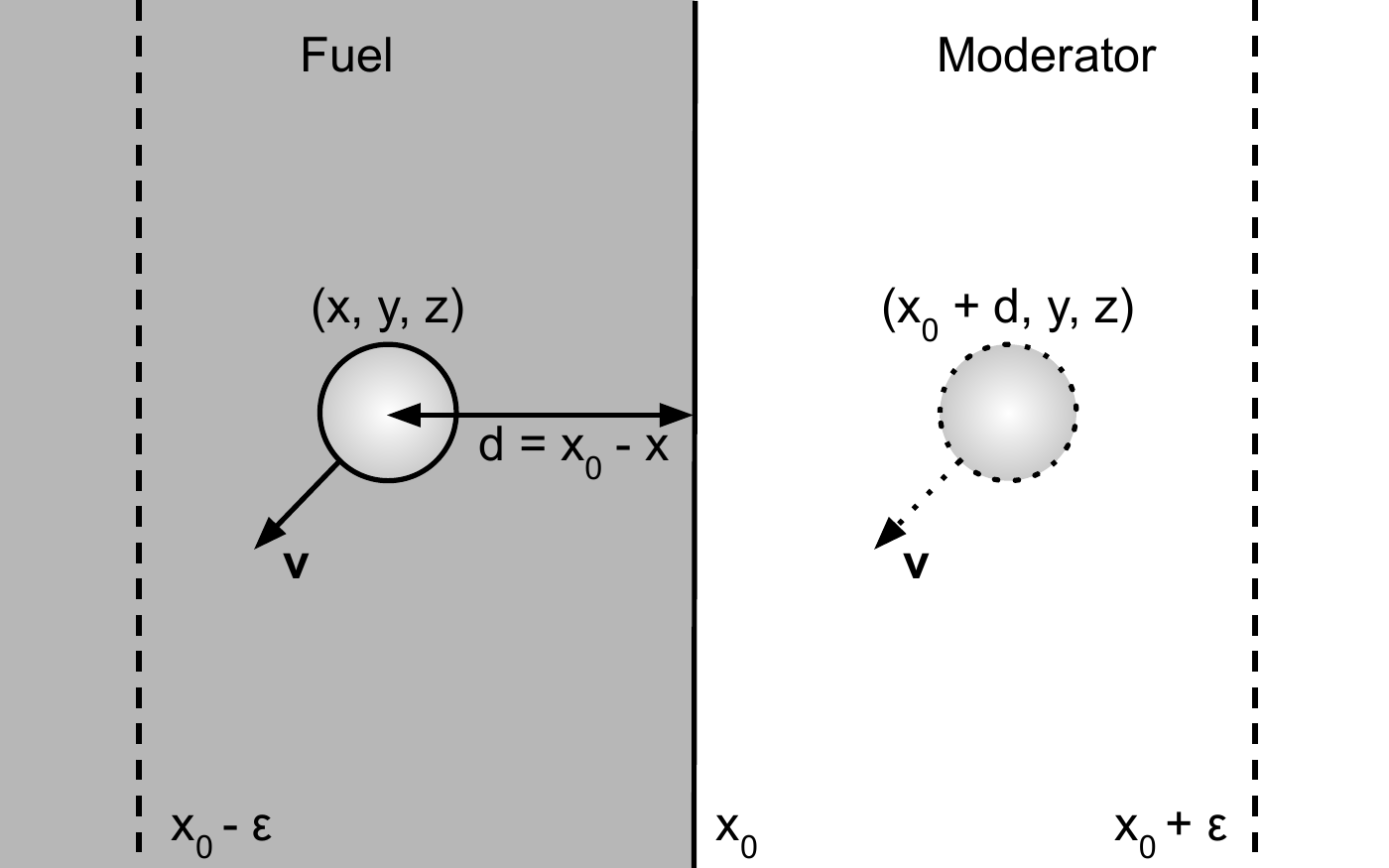}
        \caption{Depicted is the operation of copying noise particles from fission in the fuel, and placing them in the moderator. The line at $x_0$ is the boundary between the fuel and the moderator for the static problem. Dashed lines at $x_0-\varepsilon$ and $x_0+\varepsilon$ represent the minimum and maximum positions of the interface with the vibration. The particle in the fuel is the sampled fission noise particle. An exact copy of this fission noise particle (identical direction, energy, and weight) is placed at the symmetric position within the moderator, corresponding to an $x$-coordinate of $x_0+d$. The weight of the copied particle in the moderator must be multiplied by $(-1)^{n+1}$, to account for the parity of the coefficients $c_n(x,x_0)$.}
        \label{fig:fission_approx}
    \end{figure}
    
    At first glance, it would appear that Eq.~\eqref{eq:t4_vib_source} could be handled with importance sampling, as was done for cross section oscillations, where one only needs information from the material at which the collision occurred, and this is the main advantage of the proposed approximation. Sadly, further assumptions are required. As a tangible example, let us consider the case of a fuel pin vibrating in water moderator. Unlike for the case of oscillating cross sections, where all of the nuclides participating in the perturbation are present at the sampling site in the unperturbed system, in Eq.~\eqref{eq:t4_vib_source} isotopes from the fuel participate in the perturbation in the water and vice versa. It should therefore be possible to sample fission noise particles from collisions occurring in the water surrounding the vibrating pin, which would not be allowed with the importance sampling scheme from oscillations. A clever method was devised to circumvent this issue in \tripoli{}: when a fission noise particle is sampled in the fuel region, a copy is made and then translated across the boundary of the vibration, to the symmetric location relative to the vibrating interface (taking into account the parity of the given harmonic). So long as the flux is approximately constant over the distance of the translation, this is likely a reasonable approximation, allowing for the placement of fission noise source particles in the water. This approximation of the placement of fission noise particles in the moderator is depicted in Fig.~\ref{fig:fission_approx}. Otherwise, the scheme implemented in \tripoli{} samples noise source particles for vibrations in the same manner as for cross section oscillations. \tripoli{} therefore also ignores the possibility of creating scatter noise particles from isotopes present on the other side of the interface (i.e. a noise particle born from a scattering event with a fuel isotope, when having a collision in the water).
    
    \subsubsection{Exact Noise Source Sampling}
    \label{sec:exact_source}
    The noise equation is typically written using macroscopic material cross sections, yields, and energy-angle distributions. Each of the factors in the source term is actually a weighted average of the cross sections, yields and energy-angle distributions of each individual isotope present in the material. It is therefore possible to rewrite Eq.~\eqref{eq:full_perturbation_time} for mechanical vibrations as
    \begin{multline}
        \nu_\alpha(\pos,E,t)
        \Sigma_\alpha(\pos,E,t)
        f_\alpha(\pos,E\to E',\dir\to\dir',t) = \\
        \sum_k N_k(\pos,t)
        \nu_{k,\alpha}(\pos,E)
        \sigma_{k,\alpha}(\pos,E)
        f_{k,\alpha}(\pos,E\to E',\dir\to\dir')
        \text.
    \end{multline}
    Here, $N_k$ is the concentration of nuclide $k$, and $\sigma_{k,\alpha}$ is the microscopic cross section of nuclide $k$ for reaction channel $\alpha$. For an individual nuclide, the microscopic cross sections, reaction yields and energy-angle distributions are time-independent. However, the macroscopic material-averaged quantities become time-dependent through the concentration $N_k(\pos,t)$ of nuclides in the perturbed region, which is clearly a time-dependent quantity. In Monte Carlo codes, each individual nuclide is modelled explicitly, and the ensemble material averages naturally arise from neutrons undergoing collisions with all of the nuclides present in a given material. This suggests a new sampling strategy for the noise source due to mechanical vibrations. The concentration of nuclides in the perturbed region can be decomposed as
    \begin{equation}
        N_k(\pos,t) = N_k(\pos) + \delta N_k(\pos,t)
        \text,
    \end{equation}
    allowing us to write each term of the noise source as
    \begin{multline}
        \iint
        \sum_k
        \delta N_k(\pos,t)
        \nu_{k,\alpha}(\pos,E')
        \sigma_{k,\alpha}(\pos,E') \times \\
        f_{k,\alpha}(\pos,E'\to E,\dir'\to\dir)
        \varphi_c(\pos,\dir',E')
        \dd E'
        \dd\dir'
        \text.
        \label{eq:vib_mc_source_term}
    \end{multline}
    The form of $\delta N_k(\pos,t)$ depends again on the material interface behavior given in Eq.~\eqref{eq:material_time}, and therefore leads to a similar form for the Fourier transform $\delta N_k(\pos,\omega)$.
    
    To sample Eq.~\eqref{eq:vib_mc_source_term}, one might be inclined to multiply and divide by $N_k(r)$, similar to the approach taken for oscillations, so to convert Eq.~\eqref{eq:vib_mc_source_term} into a complex-weighted particle production rate. Sadly, such a strategy is not possible for the case of vibrations. Consider the case of a UO$_2$ fuel fuel pin vibrating in water. When inside the perturbed region in the water surrounding the pin, $\delta N_{^{235}\text{U}} \ne 0$, but $N_{^{235}\text{U}} = 0$, and it is therefore impossible to divide by $N_{^{235}\text{U}}(\pos)$. Our proposed solution to this problem is to define a new fictitious material which is only used for the sampling of the noise source. This material must contain all nuclides which are present in the two materials involved in the vibration, and all of their concentrations must be nonzero. The exact choice of the concentration for each isotope is arbitrary. In this work, we decided that if a nuclide is only present in the left or the right material (but not both), then its concentration in the material where it is present is taken for use in the fictitious material. If an isotope exists in both the left and right material, then the average concentration was used. The question of the optimal choice of these concentrations is left for future work. We shall denote the isotope concentrations of this fictitious material as $N^*_k(\pos)$; correspondingly, the fictitious material has a total cross section of
    \begin{equation}
        \Sigma^*_t(\pos,E) = \sum_k N^*_k(\pos)\sigma_{k,t}(E)
        \text.
    \end{equation}
    It is then possible to multiply and divide Eq.~\eqref{eq:vib_mc_source_term} by this fictitious concentration, and perform importance sampling. The collision kernel which should be used to sample the noise source is therefore
    \begin{multline}
        \frac{\Sigma^*_t(\pos,E)}{\Sigma_t(\pos,E)}
        \sum_k
            \frac{\delta N_k(\pos,\omega)}{N^*_k(\pos)}
            \frac{N^*_k(\pos)\sigma_{k,t}(E)}{\Sigma^*_t(\pos,E)}
            \times \\
            \sum_m
                \frac{\sigma_{k,m}(E)}{\sigma_{k,t}(E)}
                \nu_{k,m}(E)
                f_{k,m}(E\to E',\dir\to\dir')
        \text.
    \end{multline}
    Now, the noise source sampling strategy is as follows. During power iteration, when a neutron undergoes a collision in the region of a vibration, a new fictitious material must be constructed.
    Nuclide $k$ is sampled with probability $N^*_k(\pos)\sigma_{k,t}(E)/\Sigma^*_t(\pos,E)$, and will be used only for the sampling of the noise source particles. At this point, the sampling of the noise source is congruent to the approach used for cross section oscillations. We have effectively reduced the problem of sampling the vibration source to the problem of sampling an oscillating (space-dependent) source. If nuclide $k$ is fissile, then fission noise particles must be sampled. Next, the noise particles born from scattering are sampled. A scattering channel for the nuclide can then be sampled nominally, along with the outgoing energy and direction. All the sampled noise source particles must have their weights multiplied by the complex factor
    \begin{equation}
        \frac{\Sigma^*_t(\pos,E)\delta N_k(\pos,\omega)}{\Sigma_t(\pos,E)N^*_k(\pos)}
        \text.
    \end{equation} 
    Finally, a copy of the incident particle is also created and added to the noise source, with its weight multiplied by $\delta\Sigma_t(\pos,E,\omega)/\Sigma_t(\pos,E)$. The fictitious material is not actually needed for producing the copy noise particle, which can be handled in an identical manner to the approach taken earlier in Eq.~\eqref{eq:copy_source_definition}.
    
    It should be noted that for realistic problems vibration regions could overlap. For the case of a fuel pin with cladding, there would effectively be two material interfaces (fuel/cladding and cladding/water): if the amplitude of the vibration is larger than the thickness of the cladding, the two vibration regions will overlap. This will have the effect of creating three effective regions: one where a fuel-cladding mixture is visible, one where a fuel-cladding-water mixture is visible, and another where a cladding-water mixture is visible. Since the effects of vibrations can be combined linearly, treatment of this situation is straightforward. One just needs to ensure that all nuclides of all vibrations at a given position are included in the fictitious material, and then sum all the contributions to $\delta N_k(\pos,\omega)$ coming from the vibration of each interface. In our implementation we chose to construct the fictitious material only when it is known what vibration regions are influencing the noise sampling at the given collision site. This avoids determining all possible region overlaps and material combinations in advance, which might be rather involved depending on the geometric form of each vibration.
    
    \section{Transport of Noise Particles}
    \label{sec:mc_nn}
    
    Once the noise source has been sampled as shown in the previous section, noise particles must be transported according to the stochastic rules defined by the operator $\mathcal{B}(\omega)$, in order to solve the fixed-source problem in Eq.~\eqref{eq:noise_freq}. For both oscillations and mechanical vibrations, the structure of the noise source term is
    \begin{equation}
    -\delta {\cal B}(\omega) \varphi_c = - \sum_n \delta {\cal B}_n \varphi_c \delta(\omega- n\omega_0),
    \end{equation}
    where $\delta {\cal B}_n = \delta {\cal B}(n \omega_0)$ is the perturbation operator evaluated at frequency $n \omega_0$. (The case of oscillation is special in that the source contains only a single frequency.) Correspondingly, we therefore expect the noise field to be of the form
    \begin{equation}
    \delta \varphi({\bf r}, {\boldsymbol \Omega}, E, \omega) = \sum_{n} \delta \varphi_n({\bf r}, {\boldsymbol \Omega}, E) \delta(\omega - n \omega_0),
    \end{equation}
    where $\delta \varphi_n({\bf r}, {\boldsymbol \Omega}, E) $ is the solution of the Fourier-transformed noise equation corresponding to the noise source component at frequency $n \omega_0$, namely
    \begin{equation}
    {\cal B}_n\delta\varphi_n = -\delta {\cal B}_n \varphi_c,
    \label{eq:noise_eq_discrete_frequency_linear}
    \end{equation}
    with ${\cal B}_n = {\cal B}(n \omega_0)$ the noise operator evaluated at the discrete frequencies $n \omega_0$ of the source. This yields an infinite system of fully decoupled linear equations for the noise components $\delta\varphi_n$. For negative frequencies where $n<0$, we make use of the fact that $\delta\varphi_{-n} = \delta\varphi_n^\dagger$, the symbol $\dagger$ indicating the complex conjugate. The null frequency at $n=0$ represents the the time-averaged effect on the static flux, as the perturbation will introduce a change of reactivity into the system, which is typically neglected in the orthodox linearization approach~\cite{Zoia2021}.
    
    The operator $\delta\mathcal{B}$ is complex-valued, indicating that the noise source particles must therefore carry complex statistical weights, and each component can be positive or negative. The noise transport operator $\mathcal{B}$ is also complex-valued, and differs from the standard Boltzmann operator in three places, which therefore alter the regular flight and collision sampling procedures. We first notice that the fission production must be divided by $\keff$, which must be known before starting the noise simulation. The second modification is that delayed neutrons must have their weights multiplied by the complex yield factor $\lambda_j/(\lambda_j + i\omega)$, depending on the angular frequency $\omega$ and on the precursor family $j$. Both of these changes are minor, and straightforward to implement in most Monte Carlo codes. More problematic is the fact that $\mathcal{B}$ involves an effective total cross section which is complex:
    \begin{equation}
        \Et(\pos,E) + i\frac{\omega}{v}
        \text.
    \end{equation}
    Two different methods have been proposed to address this peculiarity. Yamamoto proposed to change the particle's weight continuously along the flight, using a complex exponential transform involving also the modification of the track-length estimator \cite{Yamamoto2013}. In this work, we use an alternative implementation proposed by Rouchon et al., which adds the real term
    \begin{equation}
        \eta\frac{\omega}{v}\delta\varphi
    \end{equation}
    to both sides of Eq.~\eqref{eq:noise_freq}, where $\eta$ is real and has the same sign as $\omega$ \cite{Rouchon2017}. Doing so effectively adds a new copy reaction with cross section $\Sigma_\omega(E) = \eta\omega/v$, having a complex yield $\nu_\omega = (\eta-i)/\eta$, and the effective total cross section becomes
    \begin{equation}
        \tilde{\Sigma}(\pos,E) = \Et(\pos,E) + \Sigma_\omega(E)
        \text,
    \end{equation}
    which is a real positive quantity.
    Flight distances of noise particles are now sampled using the effective total cross section $\tilde{\Sigma}$. At each collision, a copy of the noise particle is made with probability $\Sigma_\omega/\tilde{\Sigma}$, which has its complex weight multiplied by $\nu_\omega$. This is consistent with treatment of $\delta\Et$ given in Eq.~\eqref{eq:copy_source_definition}. In general, $\Et$ must be replaced by $\tilde{\Sigma}$ throughout the transport algorithms. Otherwise, transport algorithms are left unchanged. Variance reduction techniques such as implicit capture and roulette can still be used. In the method proposed by Rouchon et al., implicit capture is performed by multiplying the complex weight by $\Es/\tilde{\Sigma}$. For the roulette, we follow the algorithm proposed by Yamamoto, where the real and imaginary weight components undergo roulette independently \cite{Rouchon2017, Yamamoto2013}. The choice of $\eta=1$ was used, as this has been previously demonstrated to yield good performance for most frequencies of interest for reactor physics \cite{Rouchon2017}.
    
    Each component of the complex weight of noise particles can be positive or negative; correspondingly, the noise field will be estimated through the sum of these positive and negative contributions, which causes a very large variance in the scores. This phenomenon has been observed in cross section oscillation problems \cite{Vinai2021}, but is exacerbated by the case of vibrating fuel pins \cite{VinaiSubmitted2022}. On the one hand, as the noise source for cross section oscillations is constant within the vibration region, the particles of differing sign are only produced through delayed fission, which is a weak effect in most realistic systems. During noise transport, sign changes to particle weights can only occur through the copy channel, and delayed fission. Both of these reaction channels are typically weak, and this therefore leads to minimal appearances of particles of differing sign components. On the other hand, as is depicted in Sec.~\ref{sec:results_source}, and previously shown by Zoia et al.\ \cite{Zoia2021}, the noise source of a vibrating fuel rod has a positive contribution on one side, and a negative contribution on the other, leading to nearly equal quantities of positive and negative noise source particles at the beginning of the problem. The resulting noise field is effectively the difference of two nearly equal stochastic quantities, which leads to catastrophic convergence issues for the statistical uncertainty. From these preliminary investigations, it is evident that the development of specialized variance reduction techniques is required if we want to obtain usable neutron noise results via Monte Carlo methods.
    
    \section{Weight Cancellation for Variance Reduction}
    \label{sec:cancellation}
    
    Monte Carlo transport problems involving a mixture of positive and negative particles are not unique to neutron noise, and have appeared repeatedly in the literature for e.g.~performing delta tracking without a majorant cross section \cite{Carter1972,Legrady2017}, for the calculation of the higher harmonics of $k$-eigenvalue problems \cite{Booth2003,Yamamoto2009}, and for the computation of critical buckling \cite{Yamamoto2012A,Yamamoto2012B}. Based on his pioneering work on higher harmonics and critical buckling, Yamamoto's initial implementation for solving the noise equation by Monte Carlo methods utilized an approximate weight cancellation technique, as it was determined to be necessary when trying to solve the noise equation for frequencies outside of the `plateau region' $\lambda < \omega < \lambda + \beta_\text{eff}/\Lambda_\text{eff}$ \cite{Yamamoto2013}. When solving the noise equation for frequencies within the plateau region, weight cancellation was determined to be unnecessary, and was abandoned in Yamamoto's later implementation done in MCNP4C, due to the complexity of incorporating such changes in a production-level Monte Carlo code \cite{Yamamoto2018}. Furthermore, Yamamoto's initial implementation using weight cancellation was not able to estimate the variance of scores, and it was therefore impossible to determine what effect, if any, weight cancellation had on the estimation of the noise field, beside enabling the calculations to run in finite time \cite{Yamamoto2013}. The method developed by Rouchon et al.\ is able to estimate the variance of the noise field; previous results from simpler case-studies showed that convergence within the plateau region could be achieved without weight cancellation, which was therefore not implemented in the code~\cite{Rouchon2017}. Nonetheless, the convergence difficulties mentioned above for the case of vibrations has led us to begin an examination of the possible benefits of approximate or exact weight cancellation. Our recent work has demonstrated that the use of weight cancellation can provide a very large improvement in computational efficiency for the case of neutron noise induced by cross section oscillations, leading to an improvement by a factor of 90-100 on the figure of merit (FOM) when compared to the implementation in \tripoli{} \cite{BelangerPHYSOR2022}. In this section, we show how weight cancellation can be performed on neutron noise problems for the case of vibrations, and illustrate the computational gains which can be achieved through the use of the algorithm.
    
    \subsection{When to Perform Cancellation}
    
    Previous investigations in the context of weight cancellation for delta tracking methods have shown that it is most efficient to perform cancellation on the fission source \cite{Belanger2021}: fission particles are unique in that their angular distribution is almost always considered to be perfectly isotropic, and their energy is almost independent of the incident energy. This effectively reduces the dimensionality of weight cancellation, which makes exact regional cancellation methods more efficient (in terms of the amount of weight being cancelled), and reduces the bias inherent to approximate cancellation methods \cite{Belanger2021, BelangerMC2021}.
    
\begin{figure*}[t]
        \centering
        \includegraphics[width=\textwidth]{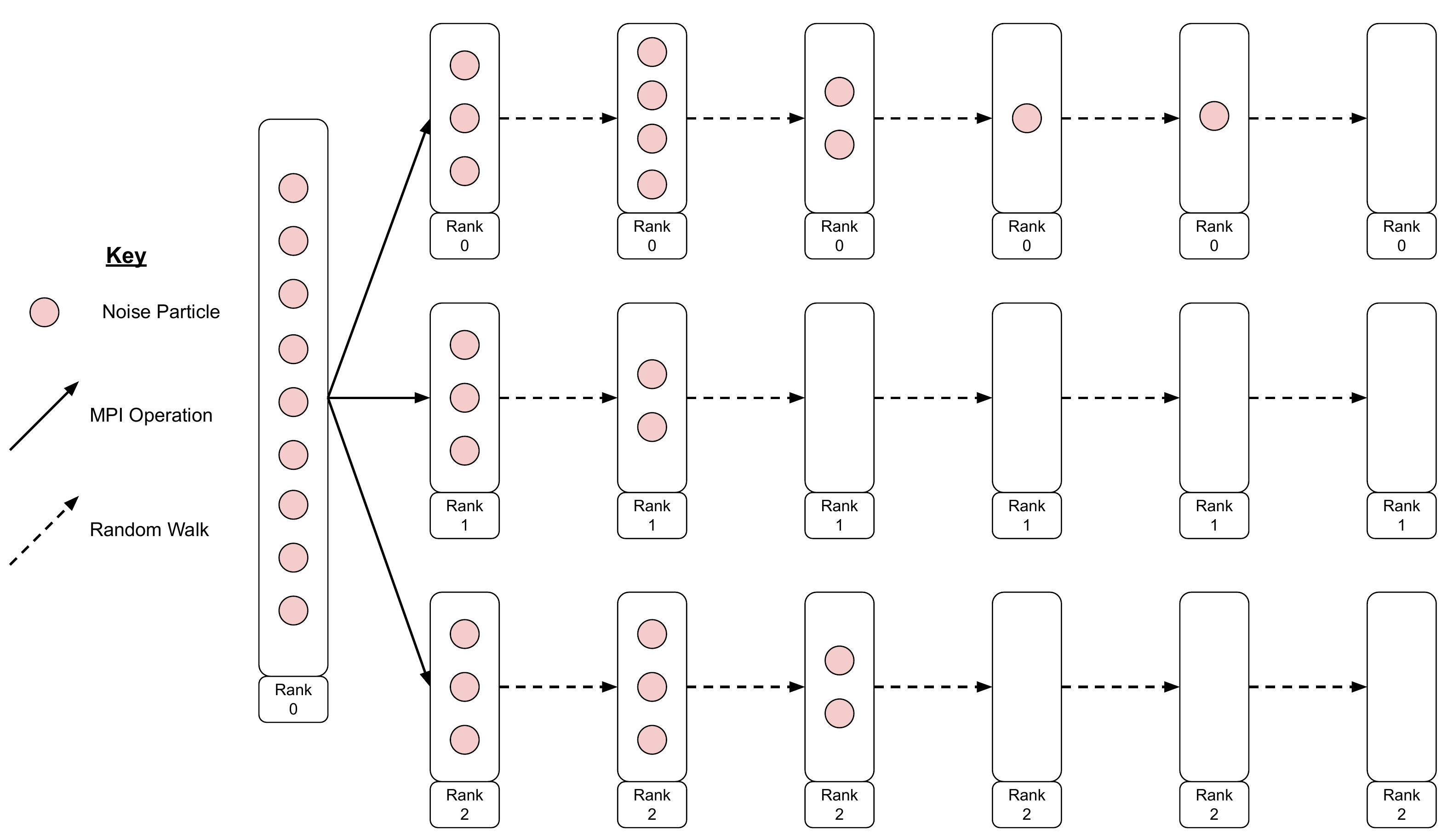}
        \caption{Standard MPI communication scheme for a single fixed-source batch.}
        \label{fig:mpi_fs_normal}
    \end{figure*}

    Most of our previous work on cancellation has been applied to eigenvalue problems solved by using the power iteration, where one naturally has access to the fission source between subsequent fission generations, providing the ideal opportunity to apply the weight cancellation algorithm \cite{Belanger2021}. However, neutron noise equations are effectively fixed-source problems, where source particles are sampled and then distributed to all the available computational nodes for the simulation. Since fixed-source problems must be always subcritical, all simulated particles eventually die: once a node receives its packet of particles to transport, it never needs to communicate with the other nodes again, until it receives the next packet of particles for the subsequent batch. Any fission particles born during a batch are typically added to the secondary particle bank, so that they can be transported later on: the node will eventually process all of the generated fission noise particles (because of sub-criticality), completing the batch. A depiction of the MPI communication scheme for this standard fixed-source algorithm is provided in Figure~\ref{fig:mpi_fs_normal}. Since all of the fission particles are just added to the bank of particles to be transported during the current batch, there is never a step in the algorithm where we have access to the entire fission source during the computation, contrary to what happens in power iteration schemes.
    
    \begin{figure*}[t]
        \centering
        \includegraphics[width=\textwidth]{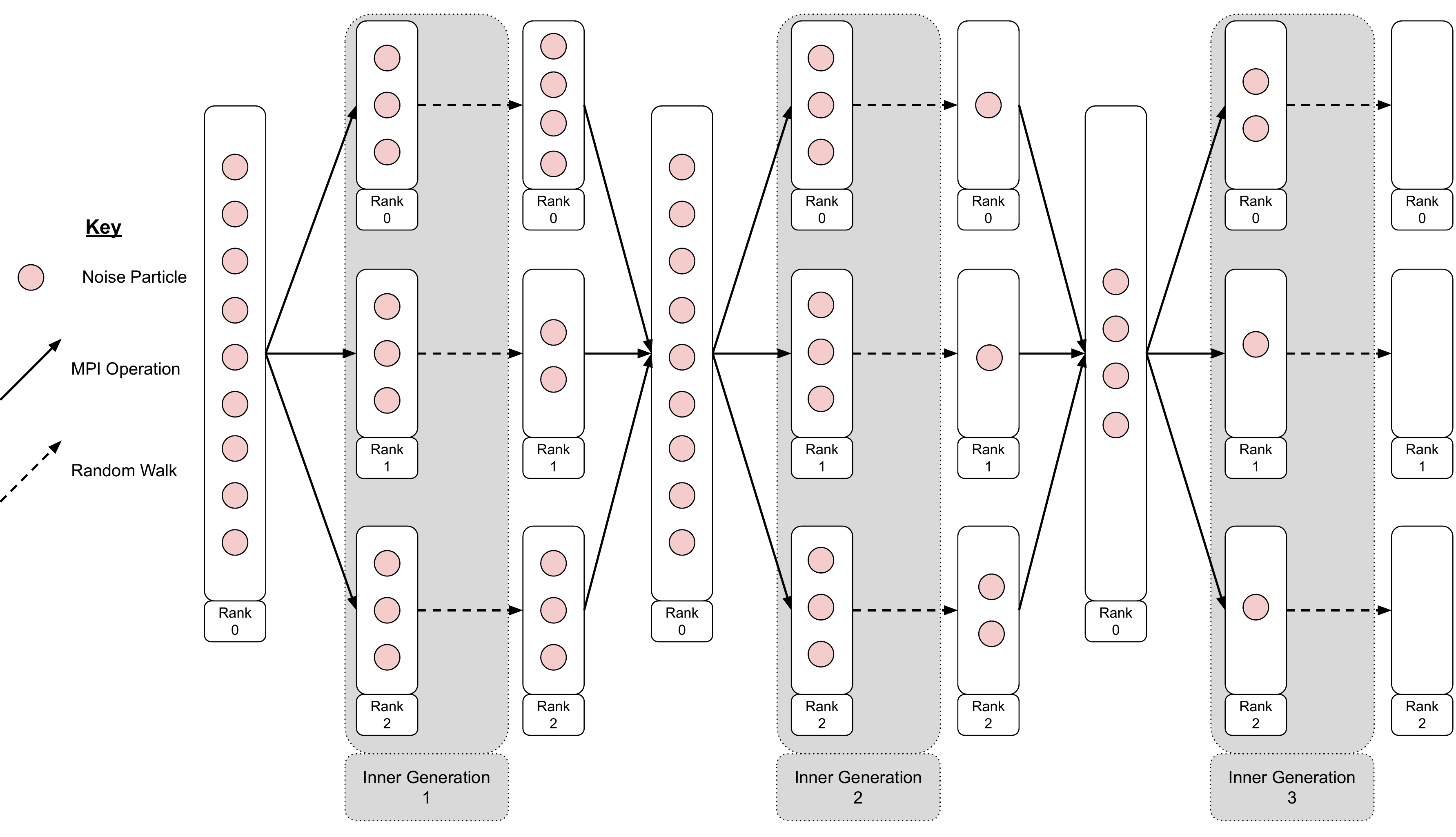}
        \caption{Proposed MPI communication scheme where a single fixed-source batch is broken into inner fission generations. While only three inner generations are depicted here to demonstrate the concept, there will generally be hundreds or thousands of inner generations per noise batch.}
        \label{fig:mpi_fs_new}
    \end{figure*}
    
    However, having access to the entire fission source is essential to maximizing the efficiency of weight cancellation: if not all of the particles participate, less weight will be canceled. We have therefore implemented a different fixed-source iteration algorithm for noise problems, in order to gain access to the entire fission source. The scheme presented here was inspired by the decomposition proposed by Yamamoto, although differing in implementation \cite{Yamamoto2013,Yamamoto2018}. For a single batch, the initial source particles are sampled, and then sent to all the available nodes. Each node will transport all of the assigned initial particles, and will store all of the produced fission particles in a separate bank. Once it has transported the initial assigned packet of particles, each node will send its bank of fission particles back to the master node. Now, with all of the fission particles on the master node, it is possible to perform weight cancellation on the entire fission source. Once the cancellation operation is complete, the new fission particles are redistributed to the nodes, where they will be transported. Any resulting fission particles will again be stored in a separate bank; once each node has finished, the fission particles are again sent to the master node, where cancellation can occur again. This practically breaks the fixed-source batch into `inner generations', as depicted in Fig.~\ref{fig:mpi_fs_new}. This MPI communication scheme is nearly identical to that used in standard power iteration implementations, with the exception that all of the particles will eventually die out, signaling the end of the fixed-source batch.
    
    \subsection{Approximate Cancellation}
    In this work, we will consider both exact and approximate weight cancellation methods. We will first examine approximate cancellation, based on the technique initially proposed by Zhang et al., which is conceptually simple, easy to implement in  a Monte Carlo code, and results in very good cancellation efficiency \cite{Zhang2016}. For this purpose, a rectilinear mesh is imposed on top of the problem domain, and the particles which will undergo cancellation are then sorted into this mesh, based on their position (and possibly energy, if desired). The average weight of all particles in each mesh element is calculated, and this average weight is then assigned to each of the particles in the element. Summing the particle weights to compute the average is what effectively leads to cancellation, as the total weight will be the sum of positive and negative components. While this approach imposes a bias on the resulting fission source, the bias can be made minimal by taking a sufficiently fine mesh. However, refining the mesh is a double-edged sword: the bias will be reduced, but the efficiency of cancellation will be reduced as well, as there will be fewer particles present in each mesh element. Despite this, our previous work has shown approximate cancellation to be much more efficient at cancelling weight than the exact version (treated in Sec.~\ref{sec:exact_cancellation}), and we have have not observed any measurable bias for a reasonably refined mesh \cite{BelangerMC2021}.
    
    \subsection{Exact Cancellation}
    \label{sec:exact_cancellation}
    Exact regional cancellation was originally proposed by Booth and Gubernatis, for use in calculating the higher harmonics of $k$-eigenvalue problems \cite{Booth2010}. The initial formulation was limited to one-dimensional configurations, but we have previously expanded this formalism to three-dimensional multi-group settings. This cancellation technique is applied on the fission source, and in practice can be used in combination with delta tracking \cite{Woodcock1965} or negative-weighted delta tracking \cite{Carter1972,Legrady2017} transport methods: we have provided a successful numerical demonstration of exact regional cancellation for a modified version of the C5G7 reactor physics benchmark \cite{Belanger2021}. Fission particles are sorted into geometric volumes, referred to as cancellation regions. Our previous work used a regular rectilinear mesh, with each mesh element acting as a cancellation region. 
    Fission particles must store the position of their parent's previous collision ($\pos'$), their parent's energy ($E'$), and their parent's penultimate direction ($\dir''$), in addition to their own phase space coordinates $(\pos,\dir,E)$. From this information, the fission density function can be calculated, which for the case of delta tracking reads
    \begin{multline}
        \zeta(\pos|\pos',\dir',E') = \\
        \frac{\mathcal{P}\left(\dirto{\pos'}{\pos}\cdot\dir'\right)
              \Sigma_f(\pos,E')
             }
             {2\pi\abs{\pos-\pos'}^2}
             e^{-\Emaj(E')\abs{\pos-\pos'}}
        \text,
    \end{multline}
    where $\Emaj(E')$ is the majorant cross section used in the delta tracking algorithm \cite{Woodcock1965,Leppanen2017,Belanger2021}. Using this fission density function, each fission particle $i$ (having a weight $w_i$) is split into two portions: a point-wise part with weight
    \begin{equation}
        w_{p,i} = \frac{\zeta(\pos|\pos'\dir'',E') - \beta_i}{\zeta(\pos|\pos'\dir'',E')}
        \text,
    \end{equation}
    and a uniform part with weight
    \begin{equation}
        w_{u,i} = \frac{\beta_i}{\zeta(\pos|\pos'\dir'',E')}
        \text.
    \end{equation}
    Here $\beta_i$ is a cancellation parameter, which can take any value, so long as it does not depend on $(\pos,\dir,E)$ \cite{Booth2010,Belanger2021,BelangerSubmitted2022}. The point-wise portion represents the piece of the fission particle which must be placed exactly at $\pos$, and the uniform portion represents the piece which can be uniformly distributed across the entire cancellation region. A thorough discussion on how to determine an appropriate value of $\beta_i$, which ultimately affects the cancellation efficiency, is provided in Refs.~\citenum{Belanger2021} and~\citenum{BelangerSubmitted2022}. For each cancellation region $\mathcal{R}$, we iterate over all of the particles in the region, summing their uniform parts
    \begin{equation}
        U_\mathcal{R} = \sum_{i\in\mathcal{R}} w_{u,i}
        \text.
    \end{equation}
    We then set all of the fission particle weights to be equal to their point-wise portion ($w_i := w_{p,i}$). Lastly, we must sample fission particles with a total weight equal to the remaining weight $U_\mathcal{R}$, which will be uniformly distributed within $\mathcal{R}$. The exact cancellation algorithm can be applied as is to neutron noise problems: the only required modification concerns the number of uniform particles to be added to each region, which for particles carrying real and imaginary statistical weights reads
    \begin{equation}
        n = \left\lfloor\text{max}\big(\abs{\Re{U_\mathcal{R}}},\abs{\Im{U_\mathcal{R}}}\big)\right\rfloor
        \text,
    \end{equation}
    each with a weight of $w = U_\mathcal{R}/n$.
    
    Compared to approximate cancellation, exact regional cancellation is much more difficult to implement in a Monte Carlo code. Additionally, it results in much less total weight cancellation as well, which can lead to only minimal performance improvements in problems which require cancellation \cite{BelangerMC2021, Belanger2021,BelangerSubmitted2022}. However, the exact regional method aligns with the Monte Carlo philosophy, in that no approximation nor bias is introduced in the sampling scheme.
    
    \section{Simulation Results}
    \label{sec:simulation_results}
    The Monte Carlo noise equation solution method devised by Rouchon et al.\ was implemented in the development version of \tripoli{} \cite{Rouchon2017, T4}. Since \tripoli{} is a mature general-purpose code with about 400 kSLOC (source lines of code), implementing all of the changes necessary to perform the exact noise source sampling technique outlined in Sec.~\ref{sec:exact_source}, or the cancellation methods described in Sec.~\ref{sec:cancellation}, would require an exceptionally large rewriting effort. Therefore, all of the methods described in the paper have been implemented in a multi-group Monte Carlo mini-app called MGMC, with approximately 13 kSLOC and thus allowing for quick implementation and testing of new transport algorithms \cite{mgmc}. MGMC is able to perform transport in general 3D geometries, composed of surface-based volumes, universes, and lattices. Shared-memory parallelism is implemented via OpenMP, and distributed-memory parallelism is implemented via MPI. Basic quantities such as flux and reaction rates can be scored using either track-length or collision estimators, across regular rectilinear meshes, and are written to binary NumPy files, for easy data processing in Python \cite{Numpy}. These features make MGMC entirely representative of a larger production-level code; therefore, it can be assumed that any improvement in calculation efficiency observed in MGMC would be similar to the improvements one could expect from implementing the same methods in a production-level code, such as \tripoli{}. Delta tracking and weight cancellation have previously been added to MGMC, along with neutron noise transport \cite{Belanger2021, BelangerPHYSOR2022}. Newly implemented for this work was the sampling the noise source corresponding to mechanical vibrations. MGMC has been released as free software, and is available under the CeCILL-v2.1 license \cite{mgmc}.
    
\begin{figure}
        \centering
        \includegraphics[width=\columnwidth]{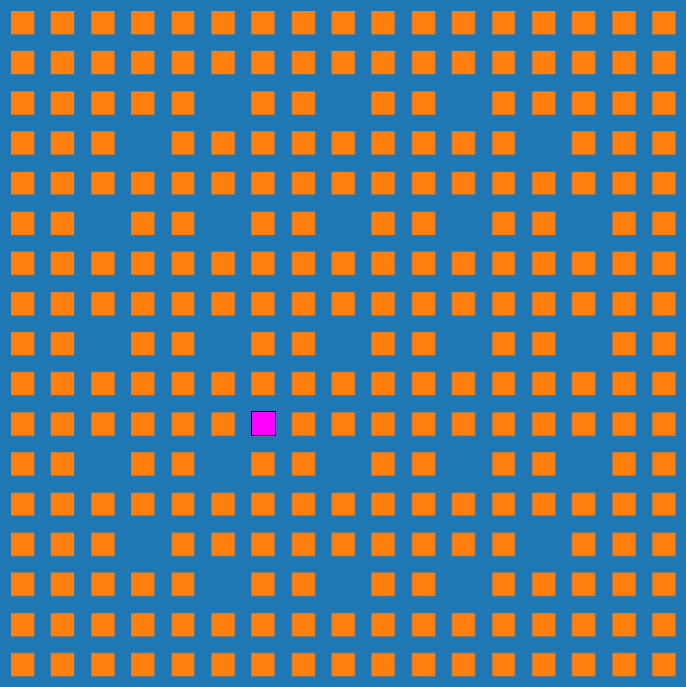}
        \caption{Depiction of the 2D reflected assembly for the neutron noise benchmark problem proposed in~\cite{Vinai2021}. The pink fuel pin (located two cells down and two cells left from the center) experiences a small sinusoidal vibration to the left and right. All other fuel pins are unperturbed.}
        \label{fig:assembly}
    \end{figure}

    In order to illustrate the novel noise source sampling and the weight cancellation methods implemented in MGMC, in this work we have selected a benchmark problem that has been previously used to compare the results of several different neutron noise solvers \cite{Vinai2021,VinaiSubmitted2022}. It is a 2D, 2 group problem, in the form of a reflected 17x17 fuel assembly, with square pins. A diagram of this system is presented in Fig.~\ref{fig:assembly}. The assembly pitch is $\SI{1.26}{\centi\meter}$, with fuel pins having a side length of $\SI{0.7314}{\centi\meter}$. A $\SI{0.08}{\centi\meter}$ water blade surrounds the entire assembly. The neutron noise is induced by a vibrating fuel pin (marked in pink in Fig.~\ref{fig:assembly}), subject to a sinusoidal displacement along the x-axis, with an angular frequency of $\omega_0=2\pi\,\si{\radian\per\second}$ and an amplitude of $\varepsilon=\SI{0.2}{\centi\meter}$, as per benchmark specifications.
    
    Since exact regional weight cancellation can only be performed with delta tracking, all simulations were run with this transport method, to ensure fairness in the comparison of simulation runtimes. All simulations began with $10^6$ particles in power iteration, distributed uniformly within the assembly, all in the first group. The first $13$ generations were discarded to allow for source convergence. Between each noise batch, three extra power iteration inactive generations were performed, to ensure proper decorrelation of the noise source between replicas. Simulations were set to run for either $10^4$ noise batches, or a maximum run time of 2 days (whichever came first). Simulations were run on a computing cluster at CEA, each run using 16 MPI processes, and 32 OpenMP threads per MPI process. The noise field was estimated with a track-length estimator in all simulations. The real and imaginary components of the field are scored in separate tallies. Breaking the scoring of complex quantities into a real and imaginary component facilitates the addition of neutron noise simulations to existing codes.
    
    For the purpose of code-to-code comparison and verification, our Monte Carlo simulations were compared to the results generated by the deterministic noise solver previously implemented in \apollo{} \cite{Rouchon2016, AP3}. The noise equation solver has been added to the IDT lattice solver of \apollo{}, which uses the method of short characteristics in conjunction with the discrete ordinates method \cite{AP3}. To accomplish this, the standard iteration loops are applied to the complex fission source and scattering source, with the addition of an iteration loop between the real and imaginary components. Thus, the standard one-group transport solver methods can be used, and one can consequently benefit from all numerical methods already implemented in \apollo{} \cite{RouchonPHYSOR2020}.
    To compare the noise source and noise field between codes, these quantities were normalized by the estimated value of the static flux at the center of the assembly in group 2, for the respective code.
    
    \subsection{Sampling of the Noise Source}\label{sec:results_source}
    
    \begin{figure*}
        \centering
        \includegraphics[width=\columnwidth]{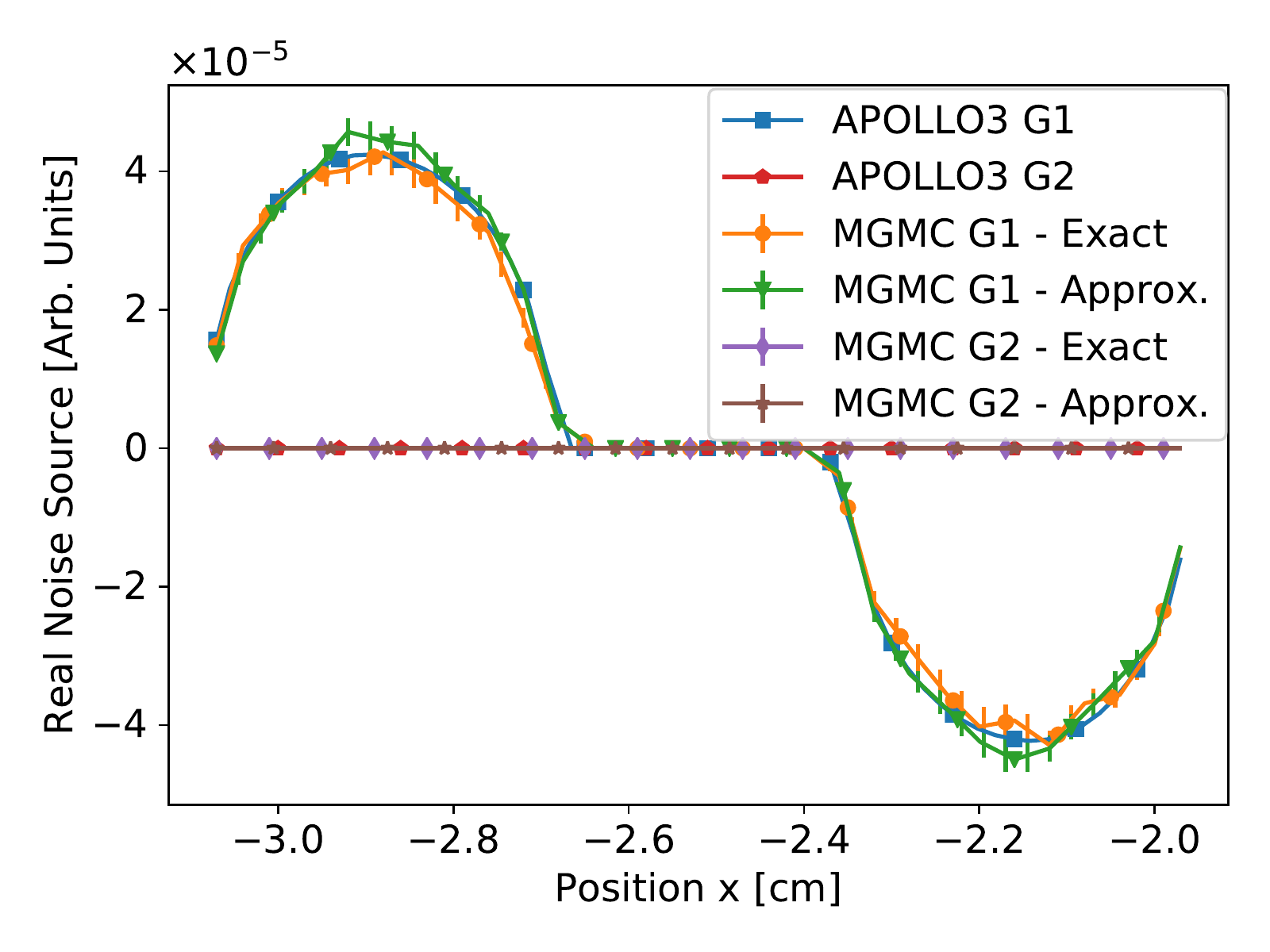}
        \includegraphics[width=\columnwidth]{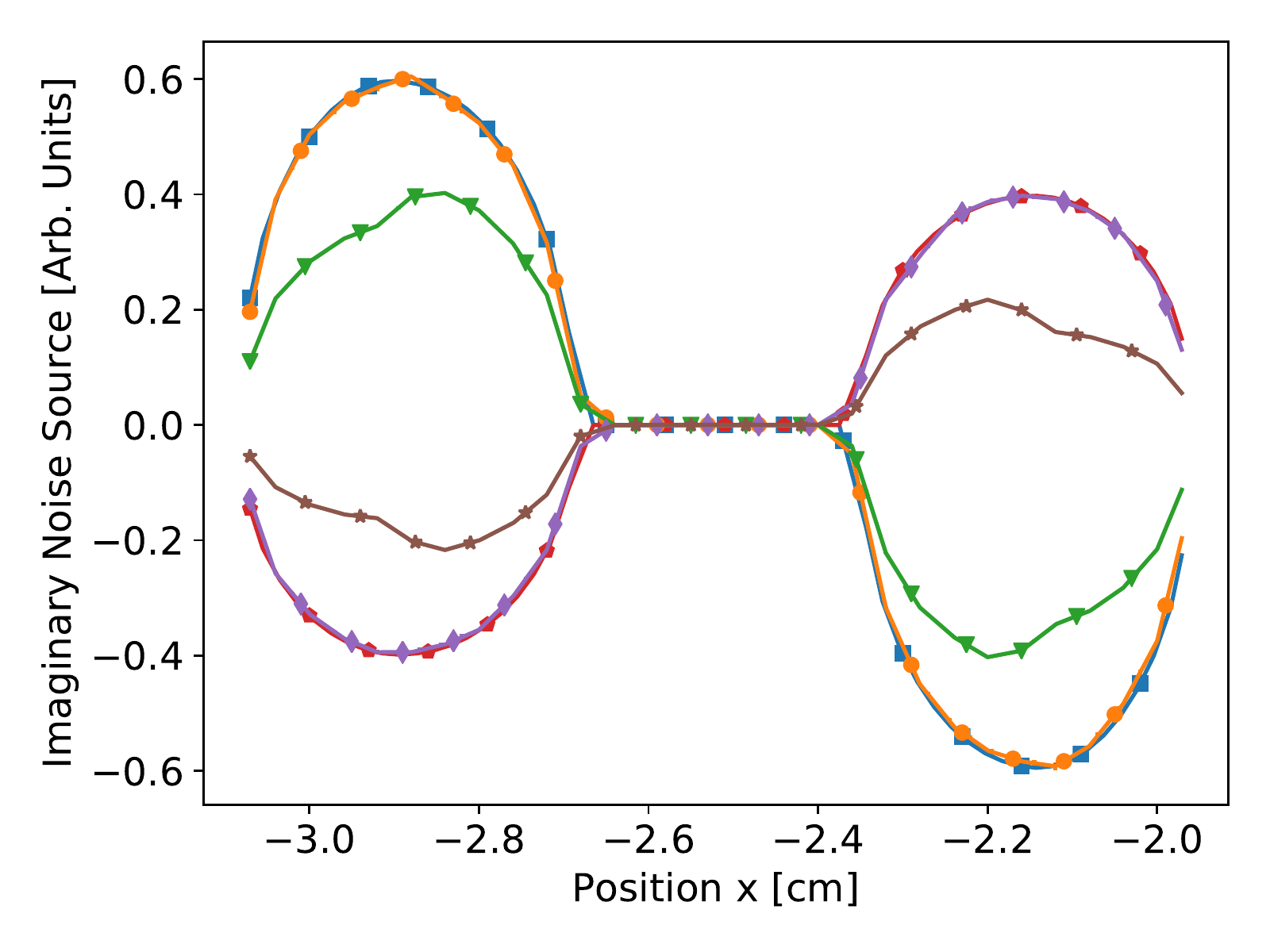}
        \caption{Real and imaginary components of the noise source, for the first harmonic.}
        \label{fig:fh_src}
    \end{figure*}
    
    We first compare the first harmonic of the noise source, sampled with the constant flux approximation and the exact method, against the noise source obtained from \apollo{}. By default, the deterministic noise solver of \apollo{} computes the harmonics of the exact noise source (Sec.~\ref{sec:exact_source}). The real and imaginary components of the source are shown in Fig.~\ref{fig:fh_src}. The real component of group 2 is zero, as this portion of the source only comes from delayed neutrons, which according to the benchmark model are all born in group 1. Immediately apparent is the fact that the amplitude of the source in the constant flux approximation is much smaller than the true source, for the imaginary component. The approximate source sampling also has an asymmetry, where the source has a slightly larger amplitude in the fuel pin than in the water. Both of these effects are primarily due to the approximate source sampling method ignoring the difference in the scattering law in the noise source. This is evident upon examination of the real component in group 1, which is purely due to delayed fission: very good agreement is observed between the approximate method and \apollo{}. Since the fission component appears to have the right amplitude and form in the real component, it can be determined that the discrepancy in the imaginary component is primarily due to the inadequate treatment of the scattering laws in the approximate method. The agreement observed in the real component also indicates that making a copy of the fission noise source particles in the fuel and moving them into the water is an adequate approximation for this system, where the flux is relatively constant over the perturbed region. Other numerical investigations concerning systems with larger flux gradients (such as the one-dimensional rod model investigated in \cite{Zoia2021}) show that this approximation might become inappropriate and lead to large differences in the resulting noise source.

    The exact noise source sampling method has excellent agreement with the \apollo{} results, for all groups and components. In the imaginary components shown in Fig.~\ref{fig:fh_src}, the exact source has the correct amplitude and shape when compared to \apollo{}. By using the same fictitious material to sample the noise source on both sides of the vibration interface, it is possible to eliminate the discontinuity which was observed on in the approximate method. This demonstrates that our proposed exact source sampling method is not just correctly treating the perturbation in the macroscopic cross sections, but also in the scattering laws.
    
    \begin{figure*}
        \centering
        \includegraphics[width=\columnwidth]{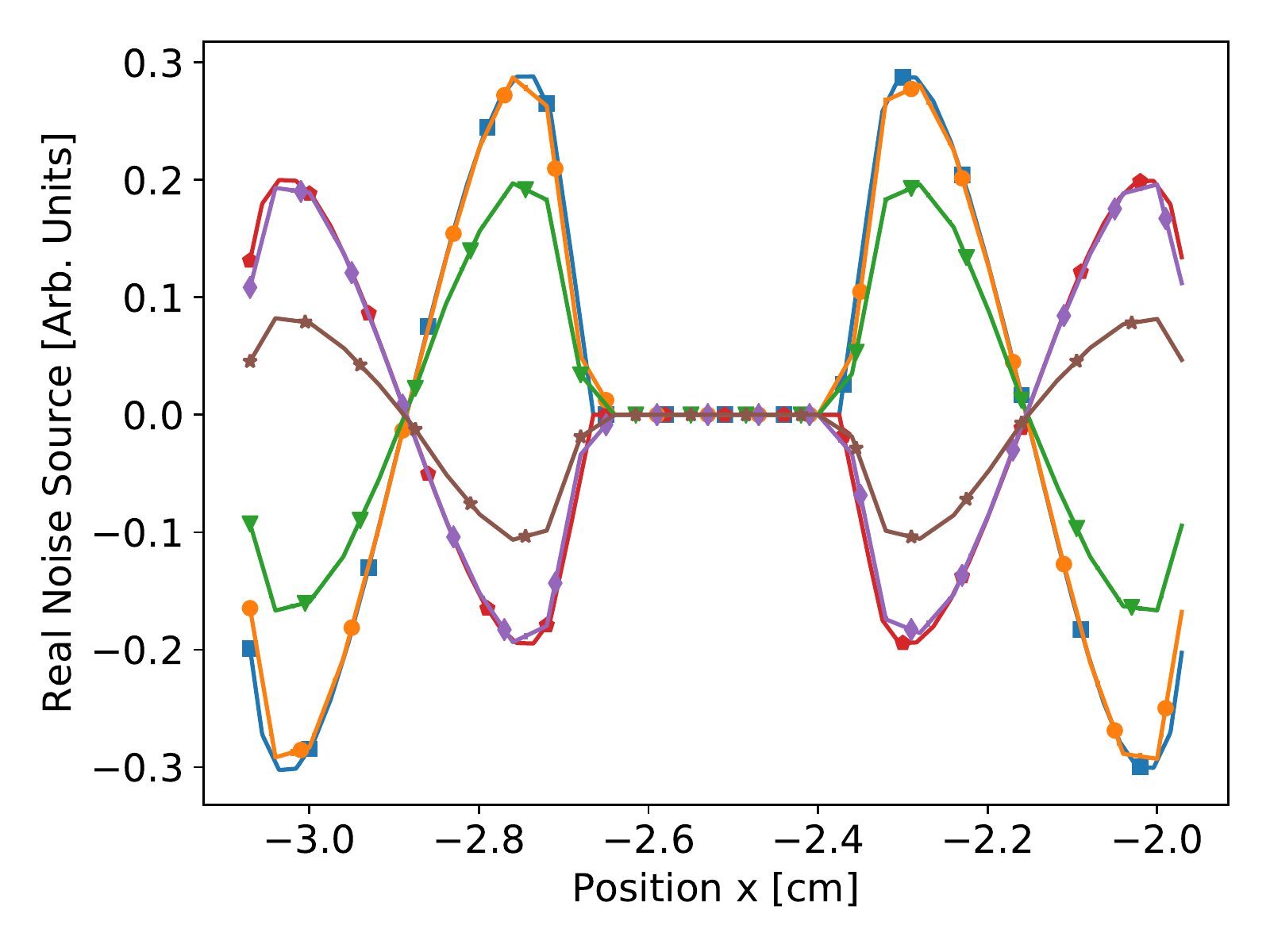}
        \includegraphics[width=\columnwidth]{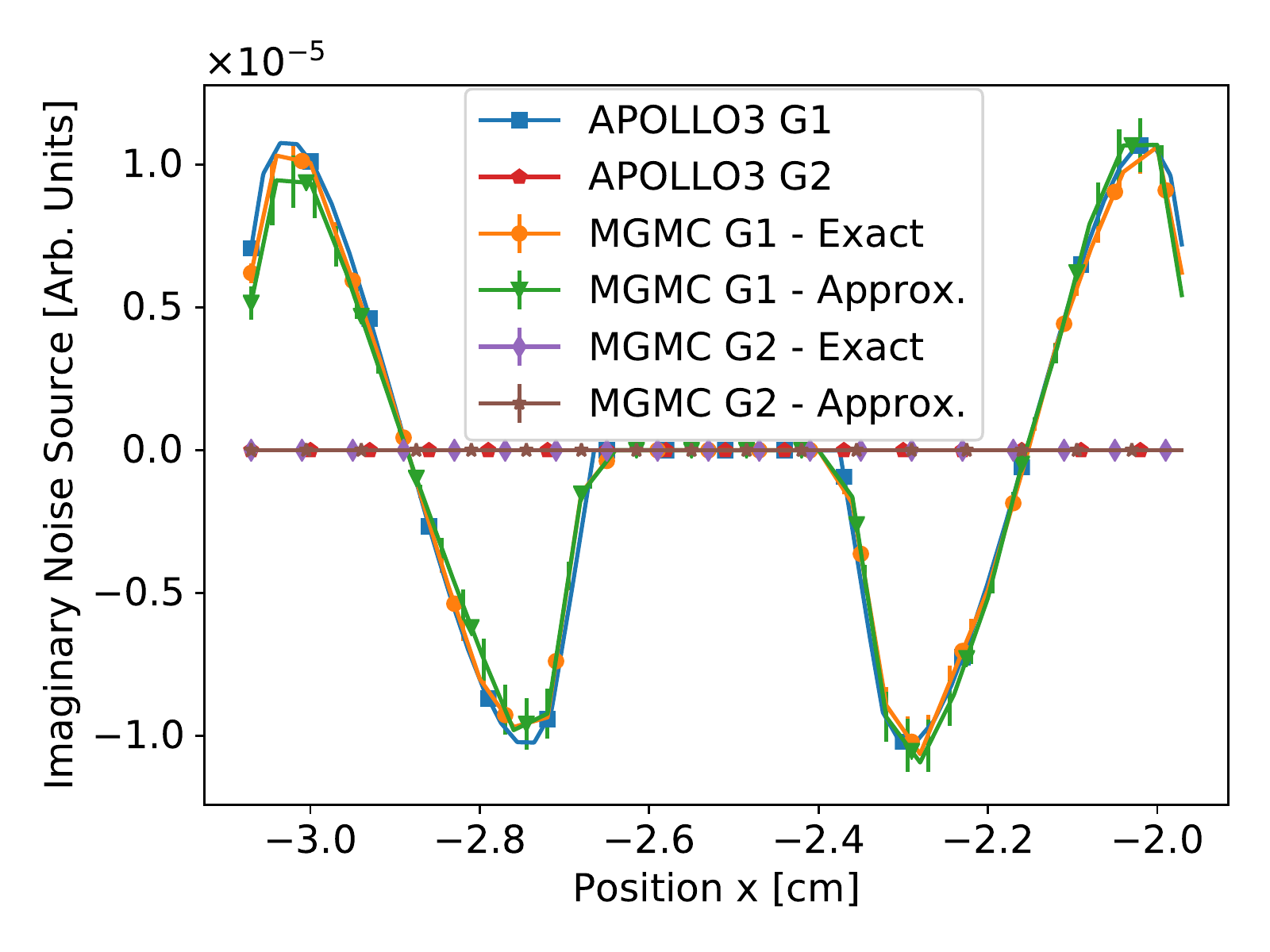}
        \caption{Real and imaginary components of the noise source, for the second harmonic.}
        \label{fig:sh_src}
    \end{figure*}
    
    Next, we look at the second harmonic of the noise source. Here, group 2 of the imaginary component is zero, and group 1 of the imaginary component comes from delayed fission, as shown in Fig.~\ref{fig:sh_src}. Most of the noise source is now in groups 1 and 2 of the real component. Again, we observe that the proposed exact noise source sampling technique is in excellent agreement with the source calculated by \apollo{}. The noise source sampled using the approximate method agrees quite well with the \apollo{} source and the exact sampling in the imaginary component, which is only due to delayed fission. Examining the real component, it is again observed that the amplitude of the source using the approximate method is smaller than the one obtained with the exact method. The asymmetry which was observed in the first harmonic is less visible in the second harmonic. In the first harmonic this asymmetry was visible exactly at the material interface. For the second harmonic, the amplitude of the real component of the first group is slightly smaller in the water than in the fuel.
    
    As the exact noise source sampling technique has now been verified, and demonstrated to have far superior agreement with \apollo{}, only the exact noise source sampling method is used for the remainder of the paper.
    
    \subsection{Application of Weight Cancellation}
    
    Next, the effects of weight cancellation on the performance of noise simulations is examined. Three different cancellation methods were used: approximate cancellation with a coarse mesh, approximate cancellation with a fine mesh, and exact cancellation. A $170\times 170$ regular rectilinear mesh was used for both the approximate coarse and exact cancellation methods. For approximate cancellation with a fine mesh, a $340\times 340$ rectilinear mesh was used instead. A simulation was also run without cancellation, and without inner generations (essentially the algorithm used in \tripoli{}), which served as a baseline for all comparisons. The resulting noise field was scored using the track-length estimator over a $138\times 138$ regular rectilinear mesh. To compare the efficiency of each simulation method, we have used the average ratio of the figure of merit (FOM) for results obtained using cancellation to results obtained using the baseline algorithm. The FOM is defined as
    \begin{equation}
        \text{FOM} = \frac{1}{T\sigma^2}
        \text,
    \end{equation}
    where $T$ is the wall-clock run time, and $\sigma^2$ is the variance of the quantity being examined. Cancellation is only applied to the noise particles, and can therefore only change the amount of time spent transporting noise particles. It will have no effect on the time spent performing power iteration to sample the noise particles to begin each noise replica. Because of this, it can be argued that the FOM should only be calculated using the time spent transporting noise particles. We have chosen to look at the FOM for both the total run time (time spent during power iteration, noise transport, and cancellation), and the noise run time (time spent during noise transport and cancellation), as we think both quantities are of interest. For each run time, the ratio for the FOM was computed in each element of the noise field scoring mesh, and then the average of that ratio was calculated independently for each component (real or imaginary) of the noise field and each energy group. As a result, for each cancellation method, 8 FOM ratios are provided, 4 calculated with the total run time, and 4 calculated with the noise run time.
    
    \subsubsection{Analysis of the First Harmonic}
    \begin{figure*}
        \centering
        \includegraphics[width=\columnwidth]{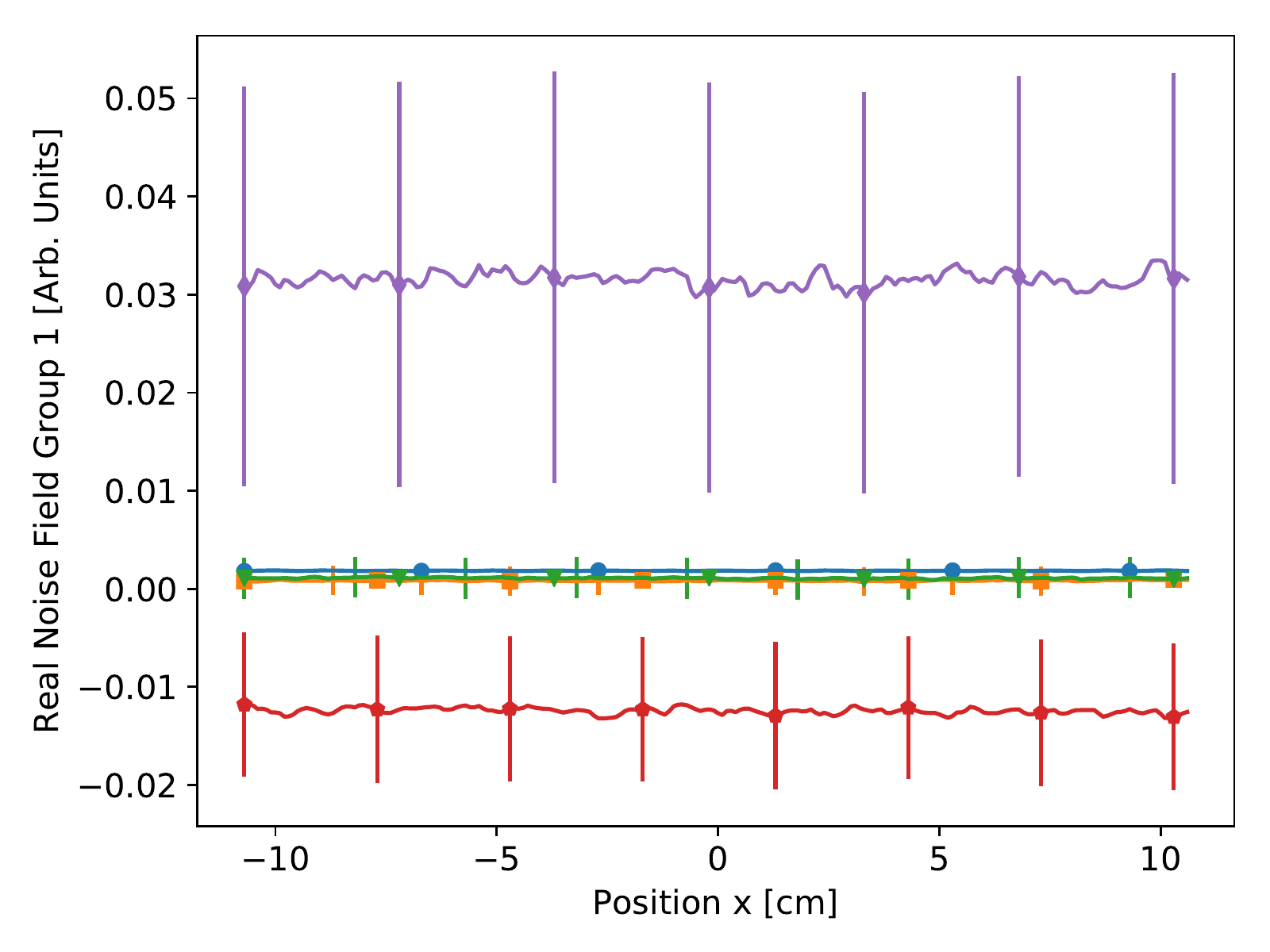}
        \includegraphics[width=\columnwidth]{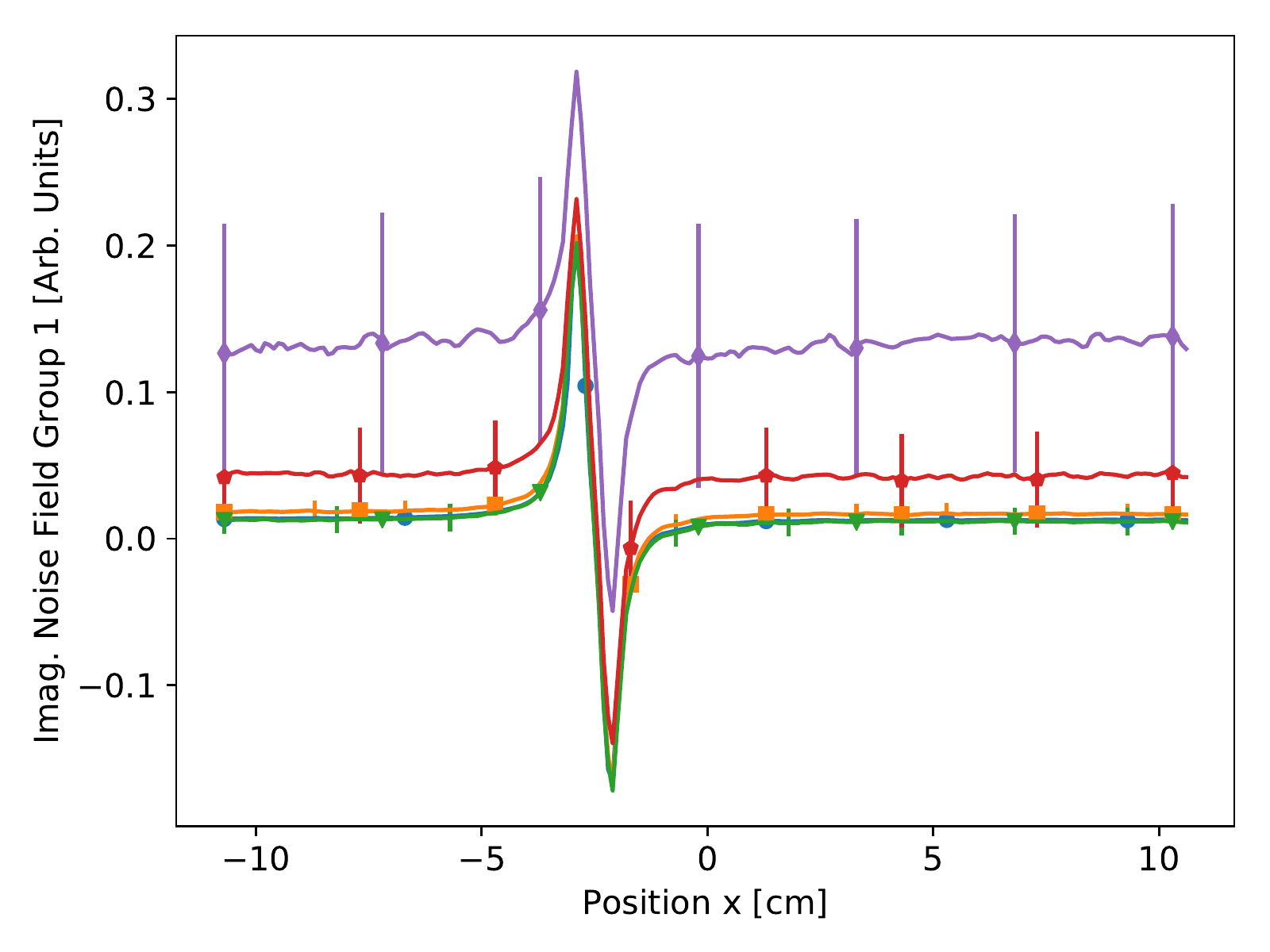}
        \\
        \includegraphics[width=\columnwidth]{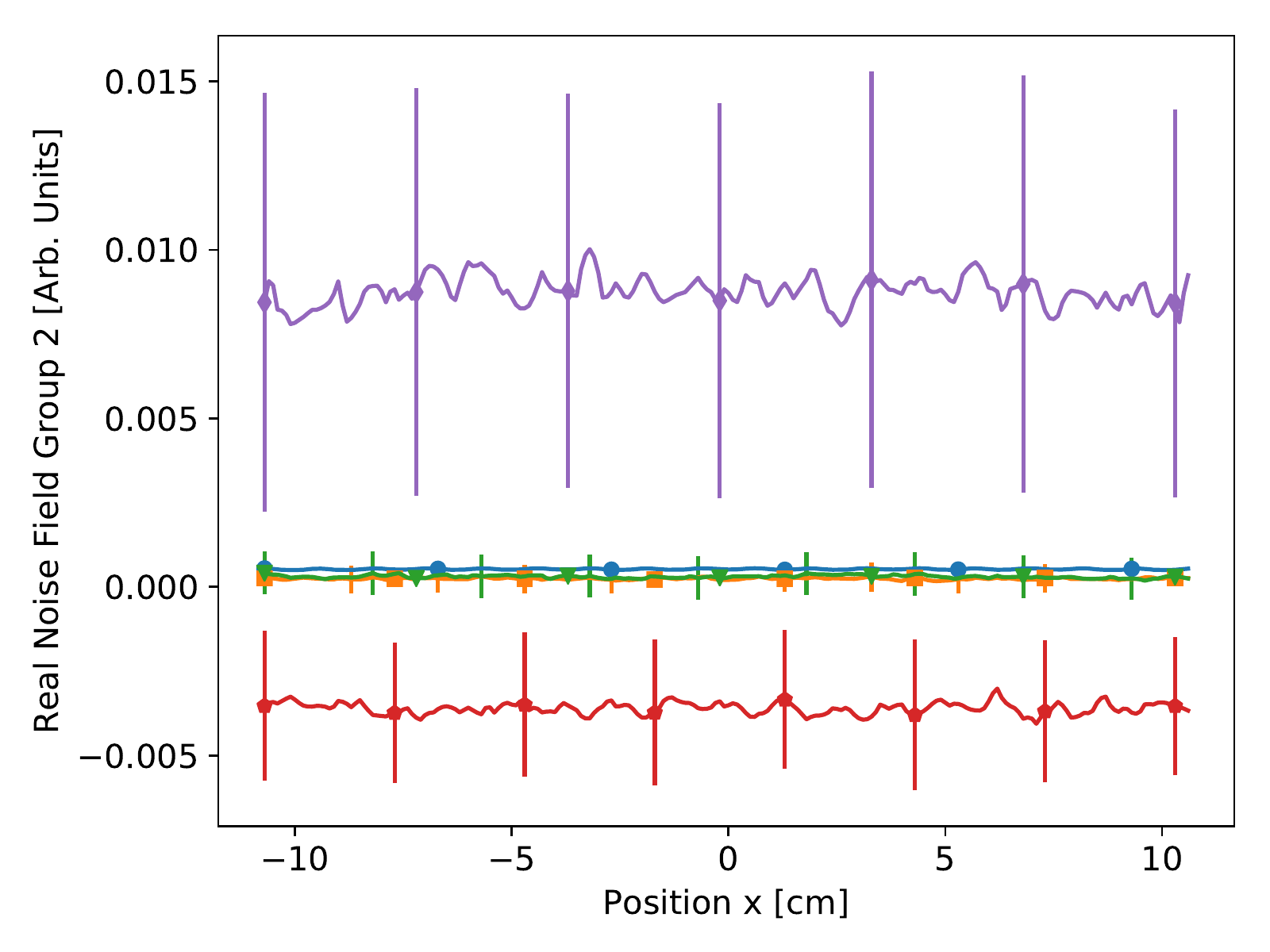}
        \includegraphics[width=\columnwidth]{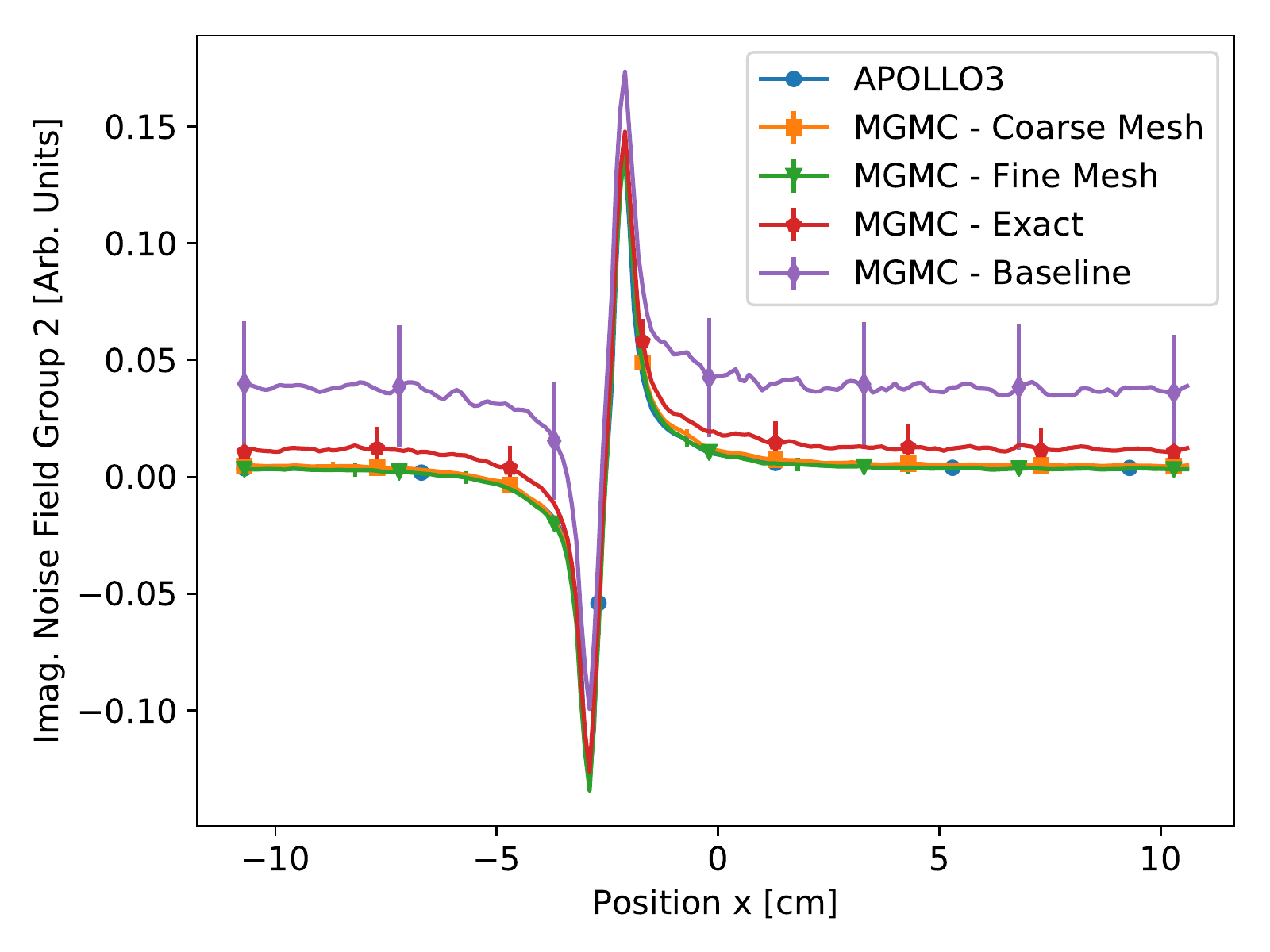}
        \caption{First harmonic of the noise field running through the center of the vibrating fuel pin, obtained with \apollo{} and MGMC using different weight cancellation methods (or no weight cancellation).}
        \label{fig:fh}
    \end{figure*}
    
    For the first harmonic of the original benchmark problem, a slice of the noise field at the axis through the middle of the perturbed fuel pin is provided in Fig.~\ref{fig:fh}. These plots compare all four the Monte Carlo simulations to the deterministic results obtained with \apollo{}. Immediately, one is drawn to the results from the baseline algorithm where no cancellation was used. Not only are the error bars quite large for both components and both groups, but it also appears as though there was a bias in the result, since this noise field systematically has a higher value than in all other simulations. This is most noticeable in the real component, and at regions far away from the perturbed pin in the imaginary component. Running the same simulation parameters with a different seed for the random number generation indicated that this behavior is not systematic, and that sometimes a similar phenomenon is observed, but with a systematically lower value than the deterministic results. This seems to suggest that the Monte Carlo results from the baseline algorithm presented here are far from converged, and would require a much larger number of noise batches to generate better statistics. Unfortunately, this was not feasible given the poor performance of this algorithm. It could also be the case that the error bars are under-estimated because of the correlations induced by power iteration; to test this hypothesis, further investigations with an increased number of decorrelation cycles would be needed. Looking at the noise field estimated with exact regional cancellation, there is certainly an observable improvement in the resulting variance, and the systematic bias observed without cancellation has been reduced. For the imaginary component, the error bars are now overlapping the deterministic results at some positions. Despite the disagreements observed far from the pin, very decent agreement is observed in the vicinity of the perturbation, where the amplitude of the noise field is largest.
    
    The two approximate cancellation methods had the best agreement with the deterministic results. While there were always visible differences in the real component, there was always agreement within the error bars. For the imaginary component, both approximate methods had nearly perfect agreement with \apollo{}, although the fine mesh variant did have slightly better agreement. These small differences could be due to the approximate nature of the cancellation technique, as refining the cancellation mesh should reduce the bias in the results. Based on the error bars and the level of agreement observed between the two approximate methods and the deterministic results, it is difficult to conclude if the slight disagreement in the coarse mesh results are actually due to the bias induced by approximate cancellation. At the observed level of agreement, one must also call into question the accuracy of the deterministic results, which have a bias induced by the geometric and angular discretization.
    
    \begin{table*}
        \centering
        \begin{tabular}{|c|c||c|c|c|c|}
            \hline
             & & \multicolumn{2}{c|}{Real} & \multicolumn{2}{c|}{Imaginary}\\
             \hline
             & & Group 1 & Group 2 & Group 1 & Group 2 \\
             \hline
             \hline
             \multirow{2}{*}{Approximate Coarse} & Total Run Time & 304 & 304 & 234 & 235 \\
             \cline{2-6}
             & Noise Run Time & 1623 & 1624 & 1252 & 1254 \\
             \hline
             \hline
             \multirow{2}{*}{Approximate Fine} & Total Run Time & 129 & 129 & 122 & 122 \\
             \cline{2-6}
             & Noise Run Time & 487 & 487 & 461 & 461 \\
             \hline
             \hline
             \multirow{2}{*}{Exact} & Total Run Time & 8 & 8 & 8 & 8 \\
             \cline{2-6}
             & Noise Run Time & 14 & 14 & 14 & 14 \\
             \hline
        \end{tabular}
        \caption{First harmonic improvement factors for the FOM of given weight cancellation methods when compared to the baseline solution strategy which does not use weight cancellation.}
        \label{tab:fh_fom_ratios}
    \end{table*}
    
    The improvements in the FOM for the simulations using cancellation, when compared to the baseline algorithm without cancellation, are presented in Tab.~\ref{tab:fh_fom_ratios}. The approximate cancellation with the coarse mesh had the best improvement in performance. Looking at the total run time, the real component was improved by a factor of about 300 and the imaginary component by a factor of about 230. If the noise run time is used, these become factors of 1620 and 1250 respectively. Approximate cancellation with the fine mesh gave an improvement of approximately 120 when looking at the total run time, and approximately 480 when looking at the noise run time. The performance gains when using the fine mesh might have been smaller due to less weight being cancelled with a finer mesh, as a finer mesh leads to fewer noise particles per mesh element. This performance penalty leads to a theoretical reduction in any bias in the results, although we were unfortunately not able to measure this phenomenon. Exact regional cancellation leads to the lowest FOM improvements, with a factor of 8 observed when using the total run time, and a factor of 14 for the noise run time.
    
    A word of caution must be added to the FOM ratios provided in Tab.~\ref{tab:fh_fom_ratios}. All of these ratios used the baseline algorithm results as a denominator. Since the FOM only considers the variance of the examined score, these reported factors do not consider the fact that the average values obtained with the baseline algorithm seemed to have a systematic bias. There is no good way to quantitatively measure the improvements in the average that cancellation provided, as demonstrated in Fig.~\ref{fig:fh}. As previously mentioned, this apparent bias could also indicate that the error bars from the results obtained with the baseline algorithm were underestimated, and the performance improvements might actually be even larger; to support this statement, one would need to perform independent replicas and estimate the `true' error bars. Additionally, the observed improvements in the FOM merit are not solely due to the application of weight cancellation. It was observed that breaking the fixed-source problem into inner generations, without applying any weight cancellation between inner generations, reduced the simulation time when compared to the baseline method. This phenomenon will be the subject of future work.
    
    \subsubsection{Analysis of the Second Harmonic}
    
    \begin{figure*}
        \centering
        \includegraphics[width=\columnwidth]{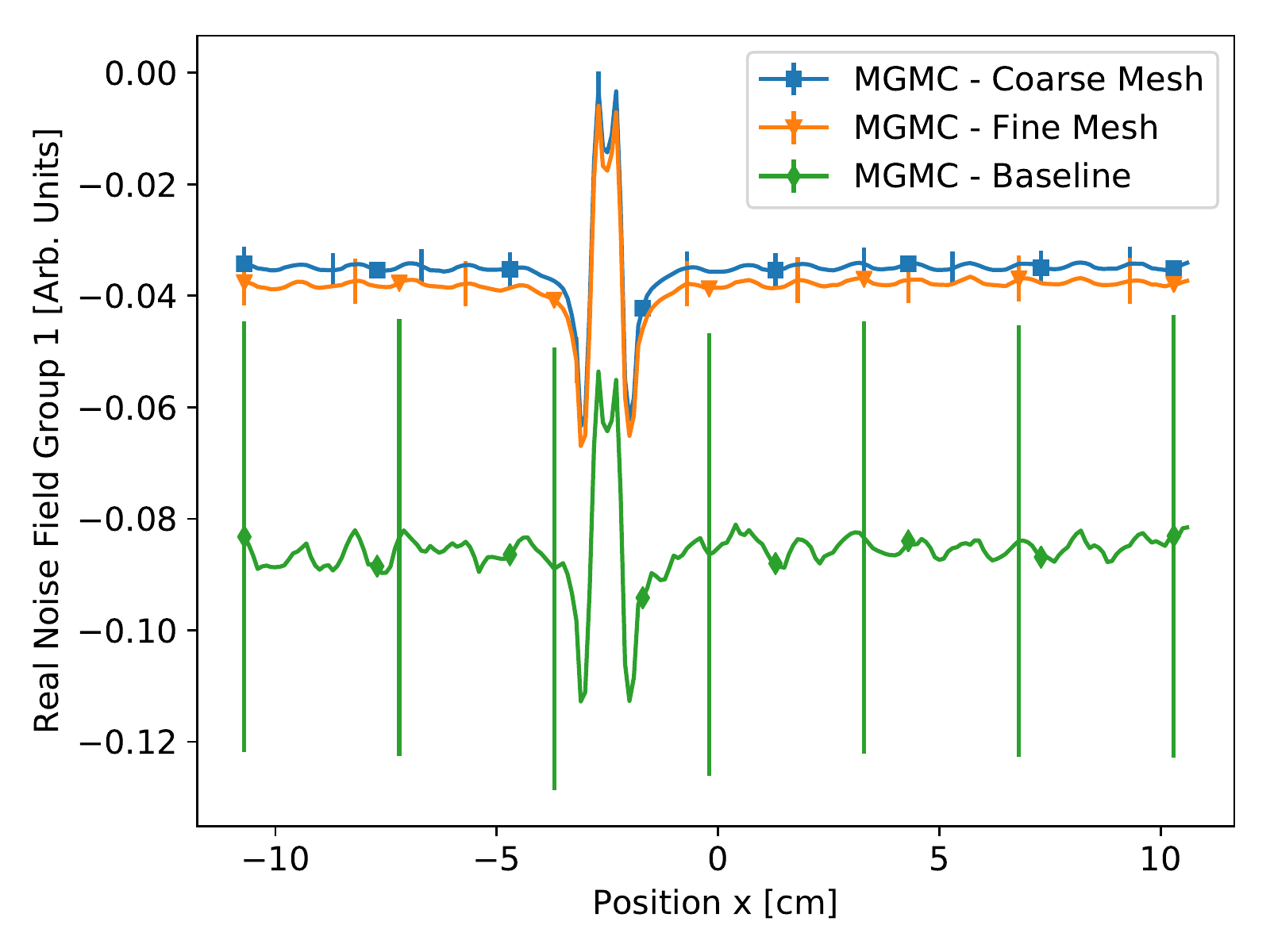}
        \includegraphics[width=\columnwidth]{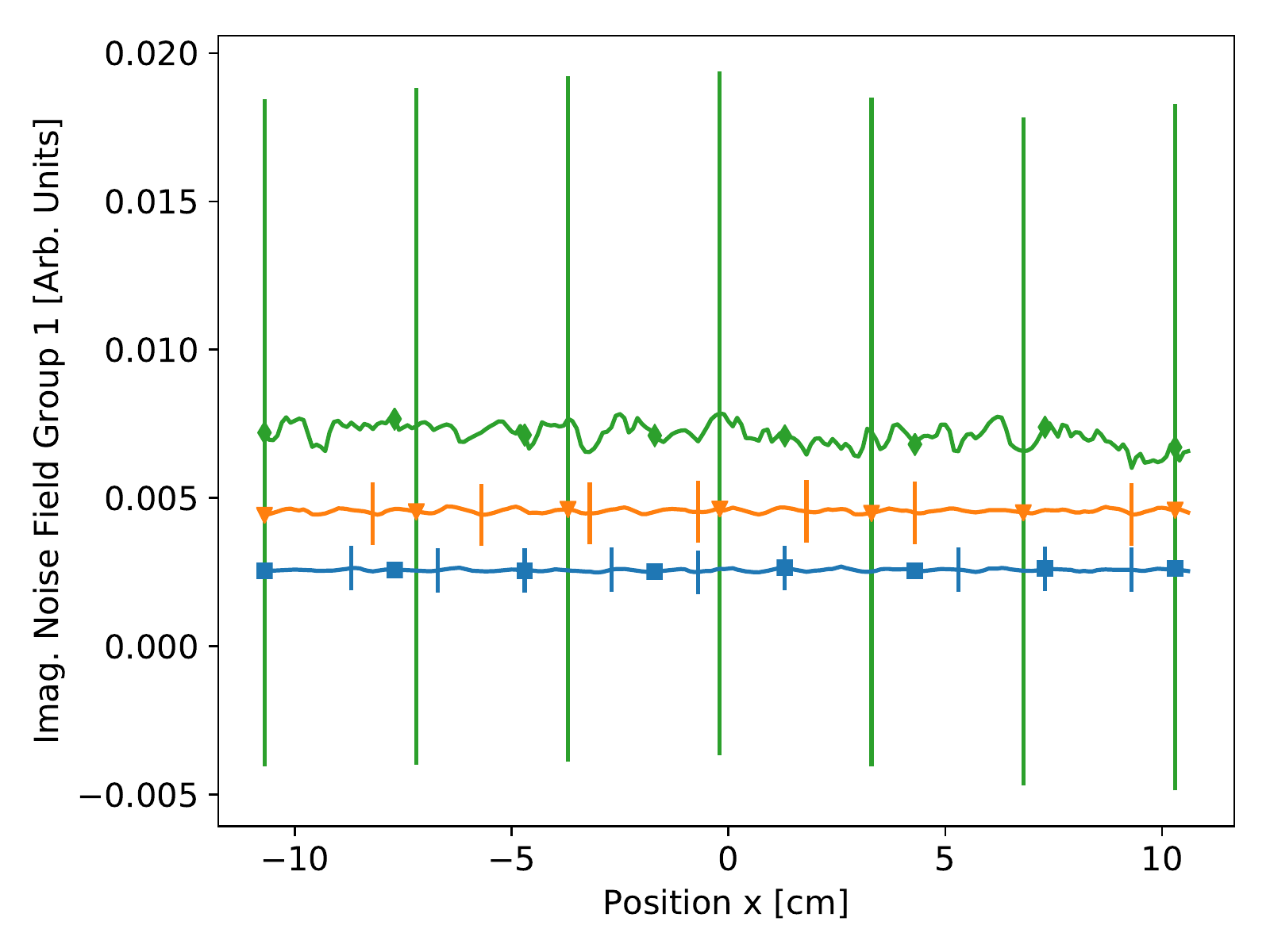}
        \\
        \includegraphics[width=\columnwidth]{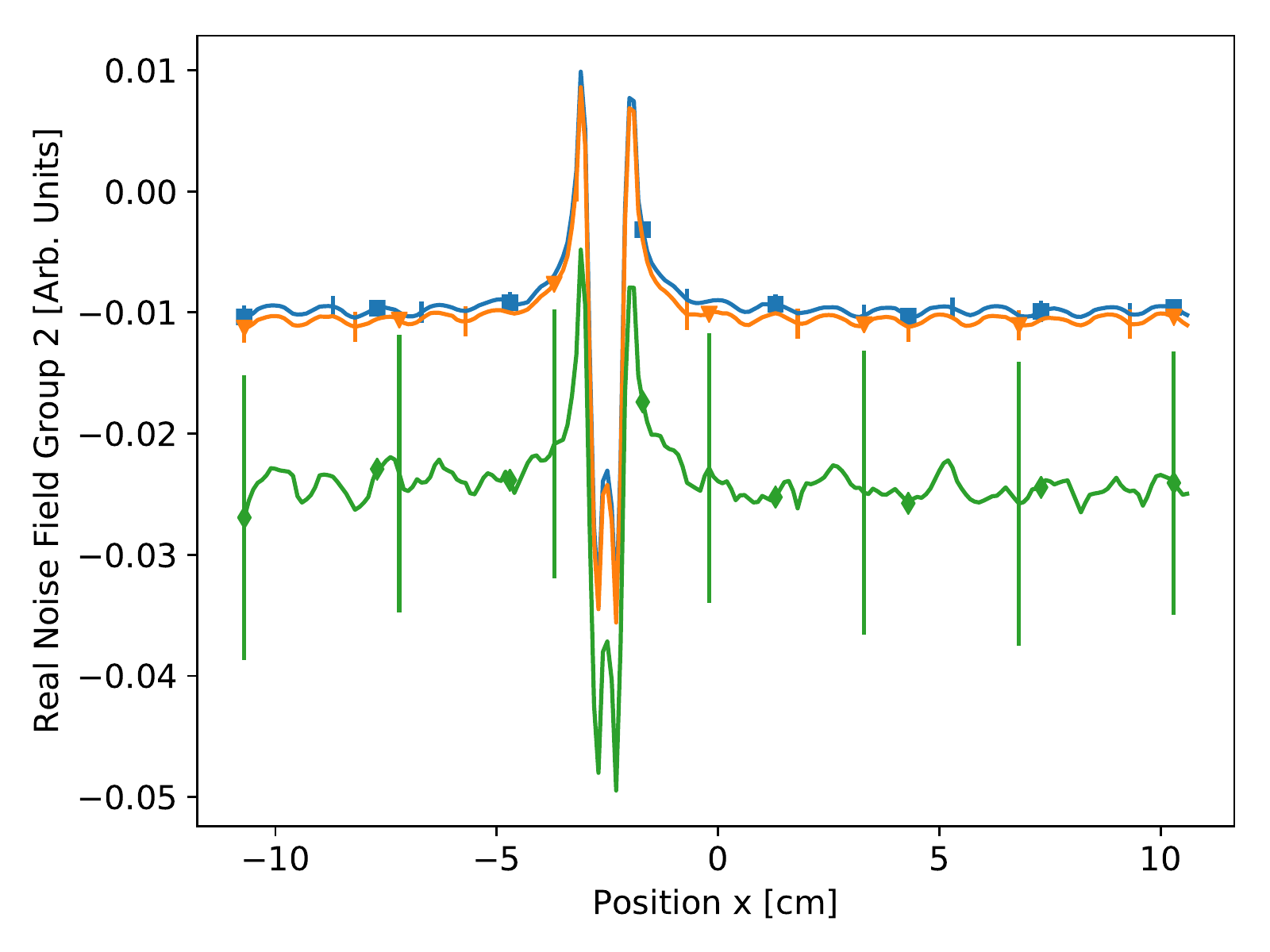}
        \includegraphics[width=\columnwidth]{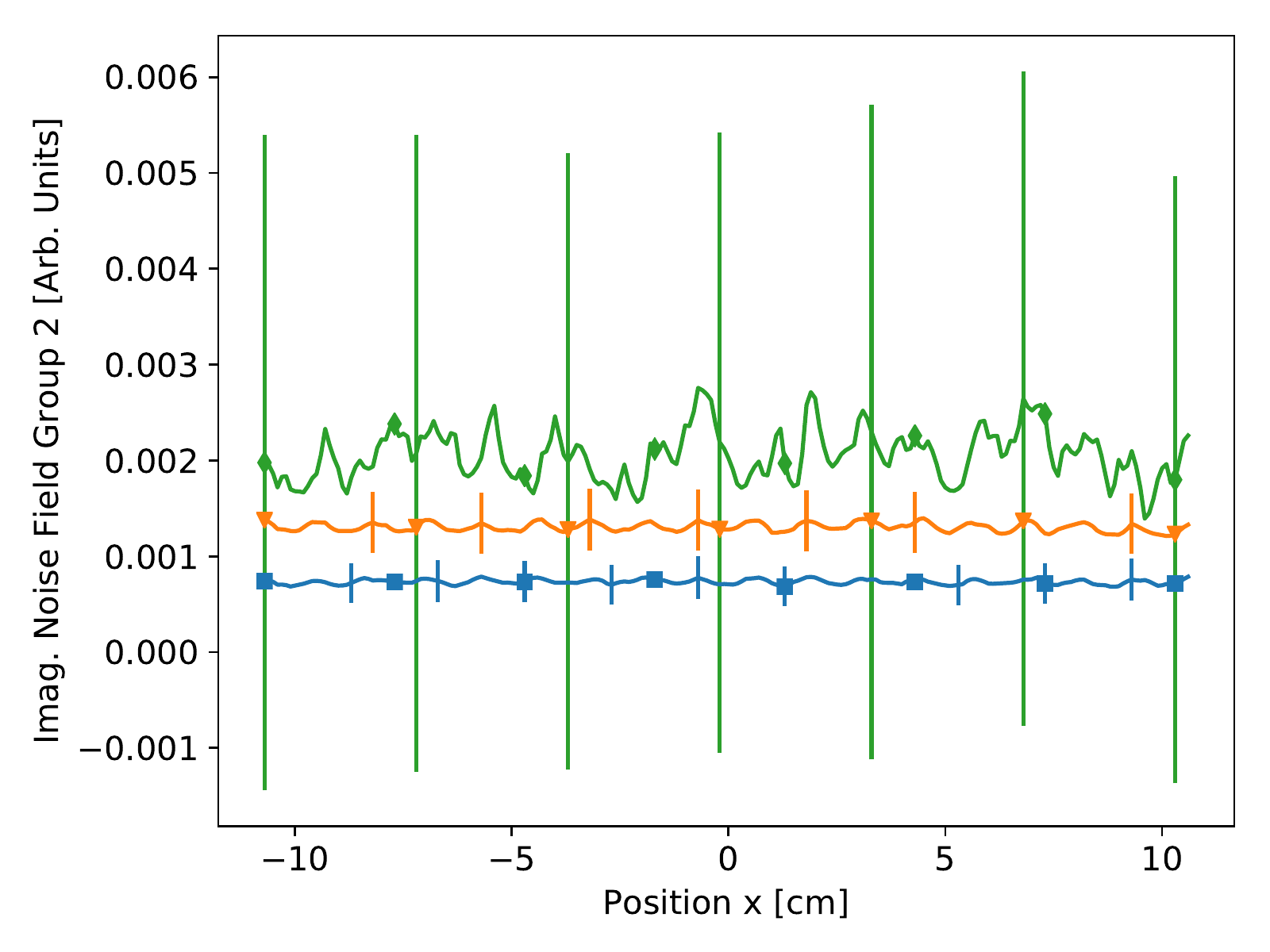}
        \caption{Second harmonic of the noise field running through the center of the vibrating fuel pin, obtained with MGMC using different approximate weight cancellation meshes, or without weight cancellation.}
        \label{fig:sh}
    \end{figure*}
    
    \begin{table*}
        \centering
        \begin{tabular}{|c|c||c|c|c|c|}
            \hline
             & & \multicolumn{2}{c|}{Real} & \multicolumn{2}{c|}{Imaginary}\\
             \hline
             & & Group 1 & Group 2 & Group 1 & Group 2 \\
             \hline
             \hline
             \multirow{2}{*}{Approximate Coarse} & Total Run Time & 244 & 244 & 344 & 344 \\
             \cline{2-6}
             & Noise Run Time & 1322 & 1323 & 1864 & 1864 \\
             \hline
             \hline
             \multirow{2}{*}{Approximate Fine} & Total Run Time & 119 & 119 & 152 & 151 \\
             \cline{2-6}
             & Noise Run Time & 446 & 446 & 570 & 569 \\
             \hline
        \end{tabular}
        \caption{Second harmonic improvement factors for the FOM of given weight cancellation methods when compared to the baseline solution strategy which does not use weight cancellation.}
        \label{tab:sh_fom_ratios}
    \end{table*}
    
    For the sake of completeness, we have then examined how our proposed approach behaves when applied to the calculation of the second harmonic of the noise field. We chose to use the two different approximate cancellation methods and the baseline method to calculate the second harmonic, to evaluate the performance gains provided by weight cancellation. It was chosen to not use exact weight cancellation here, as its performance was significantly poorer than the two approximate methods. The resulting noise fields are depicted in Fig.~\ref{fig:sh}. While the amplitude of the second harmonic in the vicinity of the vibration appears to be smaller than the amplitude observed in the first harmonic, this is not the case at positions farther from the perturbation. Far to the left or right of the vibration, the amplitude of the second harmonic is much larger than that of the first, which is coherent with previous findings by Zoia et al.~\cite{Zoia2021}. This might happen because of two possibly concurrent reasons: specific symmetries of the system could suppress the first harmonic and thus promote the second; moreover, it has been shown that the orthodox linearization of the noise equations can nonphysically amplify the second harmonic (with respect to the exact solution of the exact noise equations) while leaving the first harmonic almost unaffected. Examining the real component, it then appears as though the results from the baseline method demonstrate a large bias, and do not agree with the two other results. The estimated value of the imaginary component has much better agreement with the values of the solutions which used approximate cancellation, but there is more than $100\%$ relative error.
    
    Focusing on the two approximate weight cancellation methods, both solutions are in excellent agreement in the immediate vicinity of the vibrating pin in the real component. Moving farther away from the pin, the results obtained with the fine cancellation mesh have a slightly lower value than those obtained with the coarse mesh. It is possible that this difference is due to the finer mesh imposing less of a bias when compared to the coarse mesh. Both solutions agree within one standard deviation however, and the results would require further convergence to confirm this hypothesis. In both the real and imaginary components, the structure of the static flux is visible, which was not the case for the first harmonic results, which is coherent with the expected behaviour of local and global components of the noise field (see e.g.~\cite{pazsit_demaziere} and references therein).

    Table~\ref{tab:sh_fom_ratios} displays the FOM ratios for the two approximate cancellation methods, when compared to the baseline method. In general, the performance improvements are very similar to the values seen with the first harmonic. In the first harmonic, the imaginary component of the noise field was dominant, while real component is dominant for the second harmonic. Comparing the FOM ratio of the coarse mesh for the imaginary component of the first harmonic with the real component of the first harmonic, we see that the coarse mesh cancellation method was actually slightly more effective at estimating the noise field for the dominant component of the second harmonic than of the first. This is also true for the non-dominant component with the coarse mesh as well. Strangely, this is not true for the fine cancellation mesh, as this method had slightly poorer performance on the dominant component of the second harmonic, than on the first.
    
    \subsection{The Effects of Vibration Frequency and Amplitude}
\begin{figure*}
        \centering
        \includegraphics[width=\columnwidth]{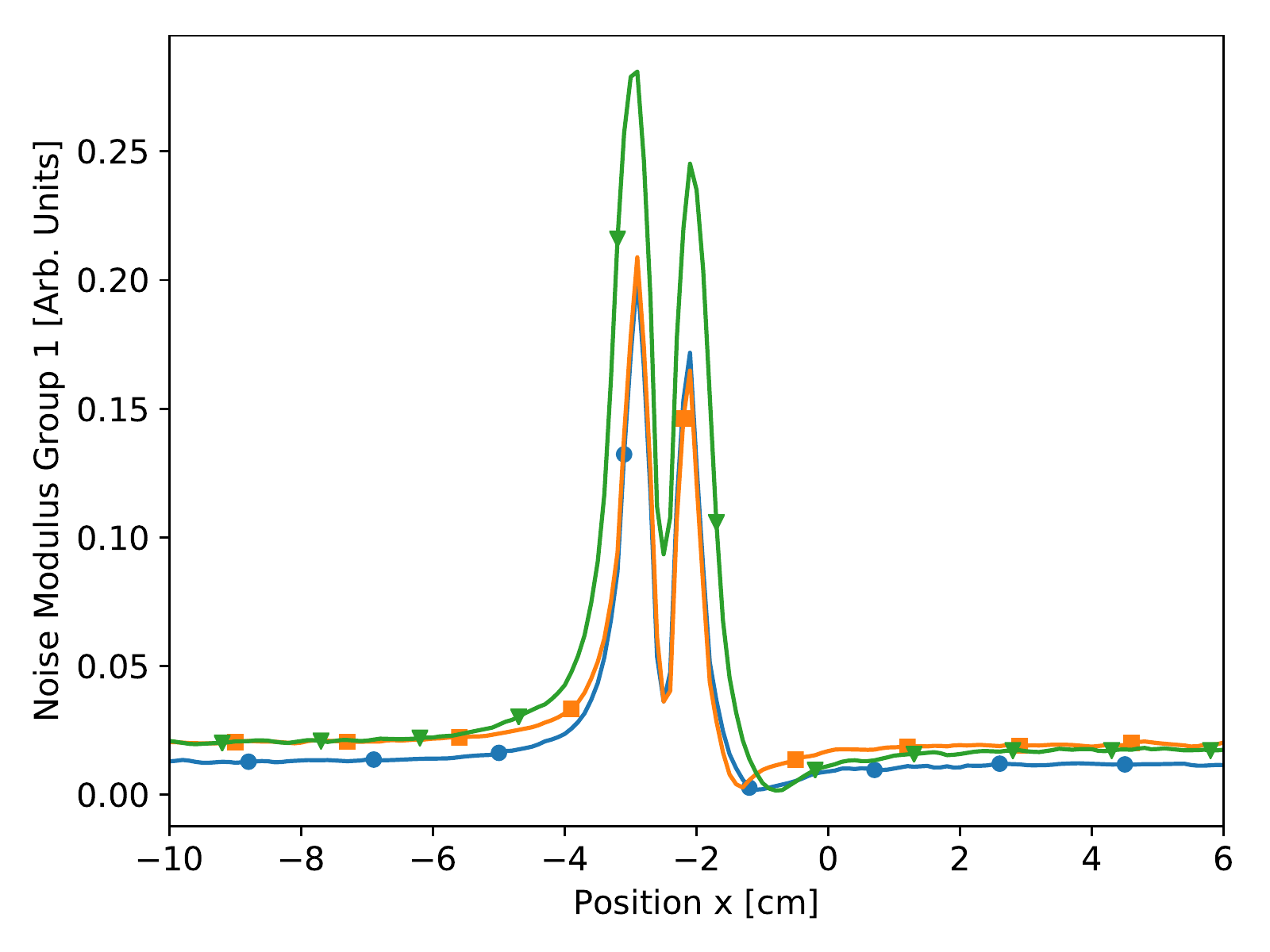}
        \includegraphics[width=\columnwidth]{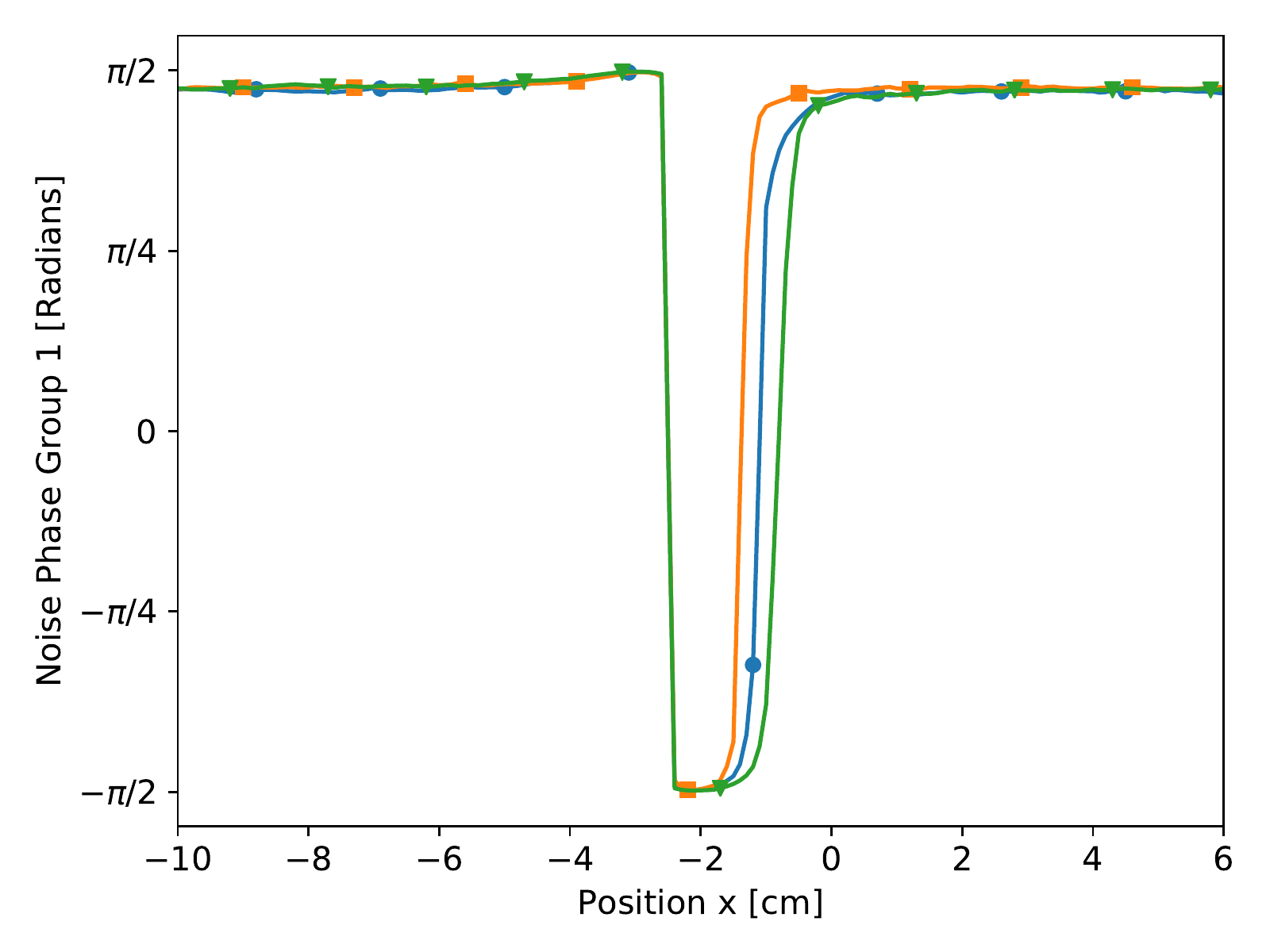}
        \\
        \includegraphics[width=\columnwidth]{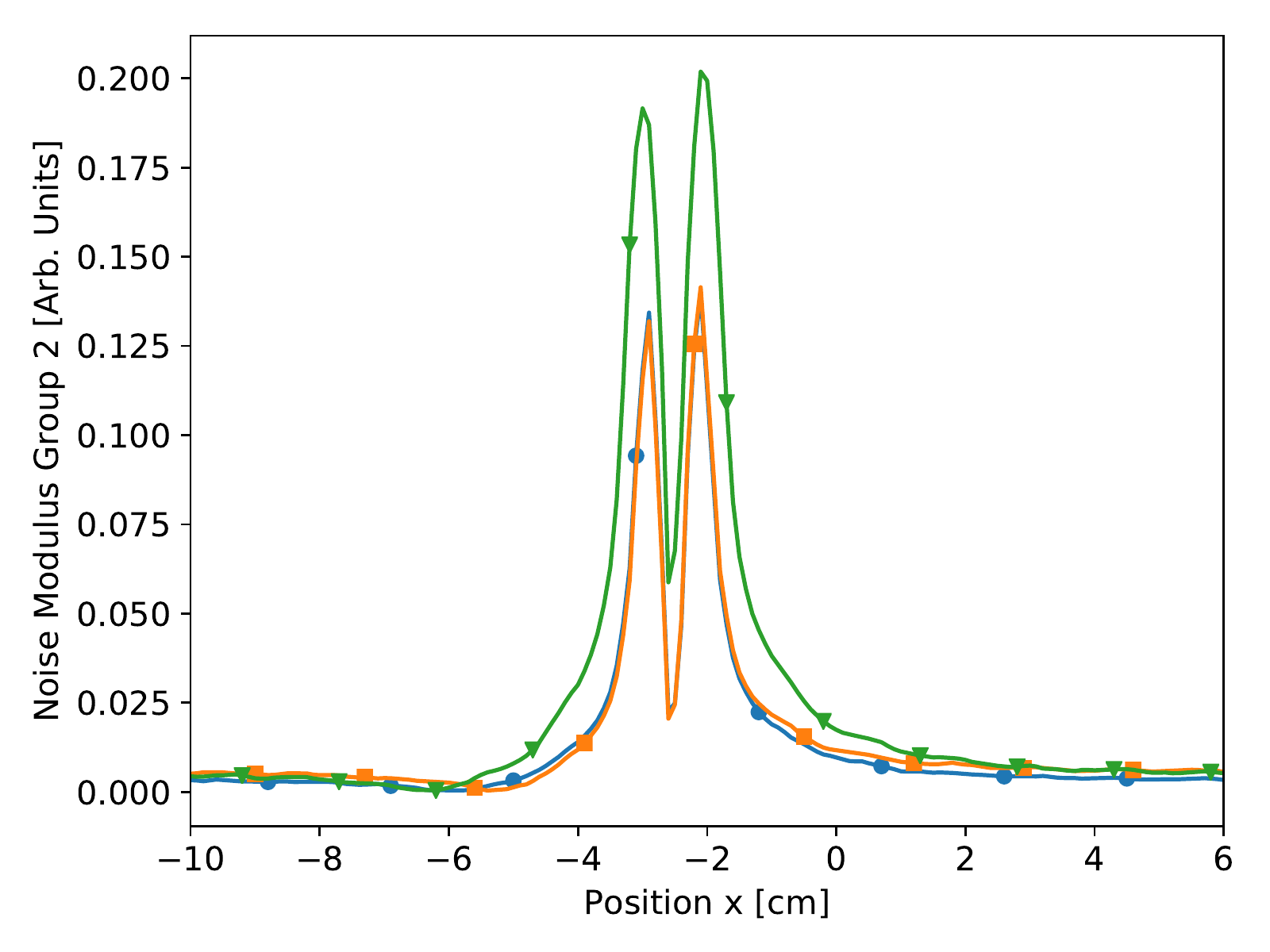}
        \includegraphics[width=\columnwidth]{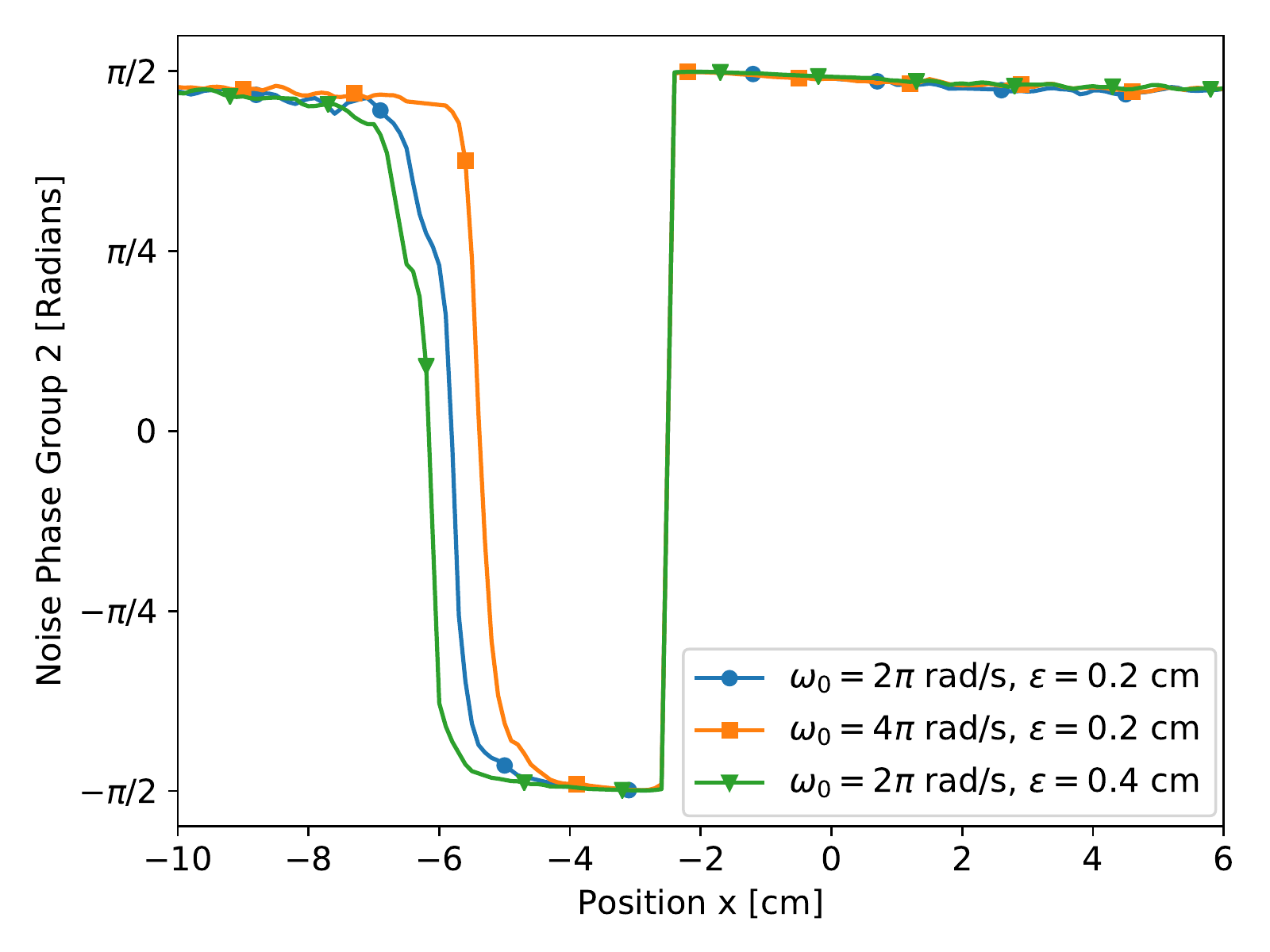}
        \caption{Modulus and phase of the noise field through the center of the vibrating fuel pin, for different vibration parameters. All results were obtained using MGMC, with approximate cancellation and the fine cancellation mesh.}
        \label{fig:comparison}
    \end{figure*}

    Given the remarkable improvements in performance that weight cancellation has enabled, it is now possible to envision the use of Monte Carlo noise simulations to perform reactor analysis and assess the impact of specific parameters on the noise field. In this section, examine the first harmonic for two altered versions of the benchmark: one where the angular frequency of the vibration of the fuel pin has been increased from $\omega_0=2\pi\,\si{\radian\per\second}$ to $\omega_0=4\pi\,\si{\radian\per\second}$ (without changing the amplitude), and a second one where the vibration amplitude has been increased from $\varepsilon=\SI{0.2}{\centi\meter}$ to $\varepsilon=\SI{0.4}{\centi\meter}$ (without changing the frequency). The resulting noise amplitude and phase for these two problems (in addition to modulus and phase of the original benchmark parameters) are presented in Fig.~\ref{fig:comparison}. Approximate cancellation with the fine mesh was used to obtain these results. In the previous sections, we examined the complex noise field $\delta \varphi$ because this is the quantity which is estimated in the Monte Carlo simulation, and the associated error bars are available, permitting a comparison in performance with the FOM. Reactor physicists, however, are typically more interested in the modulus ($\abs{\delta\varphi}$) and phase ($\atantwo\left(\Im{\delta\varphi},\Re{\delta\varphi}\right)$), which can be calculated from the complex field. Despite having error bars for the complex field, it was not possible to exhibit error bars for the modulus and phase, as the covariance between the real and imaginary part was not scored. Due to this limitation, no error bars are given in Fig.~\ref{fig:comparison}.
    
    First, we will consider the effects of the proposed variations to the vibration parameters on the noise modulus. From Fig.~\ref{fig:comparison}, we first note that in the immediate vicinity of the perturbation, changing only the frequency seems to have little impact on the modulus. Increasing the amplitude of the vibration however, leads to an increase in magnitude and width of the peaks. One would certainly expect a widening of the peak in this case, as the area of the perturbation has been increased. With a larger perturbation region, more noise source particles will be sampled in the vicinity of each interface, contributing to the larger magnitude. Farther away from the perturbed pin, we see that both the $\omega_0=4\pi\,\si{\radian\per\second}$ and the $\varepsilon=\SI{0.4}{\centi\meter}$ cases have a slightly larger amplitude, although this effect is slightly less apparent in group 2. These results could indicate that for this system the global component of the noise is more sensitive to the frequency of the vibration, while the local component is more sensitive to the amplitude of the vibration, in agreement with the spectral analysis previously performed by Rouchon \cite{Rouchon2016}.

    Examining the noise phase, it appears as though augmenting the perturbation (either in frequency or amplitude) has little effect on the depth of the observed well. Changing the vibration parameters did seem to change the width of the phase well in both cases, and this effect is apparent in both energy groups. The increased frequency reduced the well width, while the increased amplitude enlarged it. Curiously, the observed changes in the well width only ever occur on $+x$ side of the well in the first group, and on the $-x$ side of the well for the second group. A physical explanation for these observed phenomena would require further investigations.  
    
    \section{Conclusions}
    \label{sec:conclusions}
    In this work, we have considerably extended our previous findings concerning the use of Monte Carlo methods for neutron noise simulations. The contributions presented in this paper are twofold. First, we have proposed a novel method to sample in an exact manner the noise source particles produced from mechanical vibrations within a system. Contrary to the previous implementation in \tripoli{}, which for the sake of simplicity utilized many approximations, we have shown that the noise source due to mechanical vibrations can be represented without any approximation by mapping it to a perturbation of the isotopic concentrations. For this purpose, a fictitious material must be defined, containing the union of the nuclides on either side of the vibrating interface. A nuclide is then randomly chosen from this fictitious material, which allows sampling the noise source particles in a manner which is nearly identical to the method used in the much simpler case of cross section oscillations. This new method was implemented in the multi-group Monte Carlo mini-app MGMC and verified against the noise source calculated with the deterministic noise solver in \apollo{}: the two noise sources were in excellent agreement.
    
    Second, we have applied weight cancellation methods as a variance reduction technique for neutron noise simulations. For the case of mechanical vibrations, the noise source contains nearly equal positive and negative contributions in both the real and imaginary components, which makes it very difficult to estimate the resulting noise field, because of very large variances in scores. Preliminary investigations on the case of cross section oscillations have shown that these issues can be overcome using weight cancellation methods~\cite{BelangerPHYSOR2022}. In view of probing the effects of weight cancellation for the more involved case of mechanical vibrations, we have implemented a new Monte Carlo noise solver in MGMC. We have started from the Monte Carlo neutron noise method which was previously demonstrated in \tripoli{}, as it is able to produce error bars for the noise field. This solution scheme was then modified, breaking the fixed-source noise batches into inner fission generations, in order to accommodate weight cancellation. To test our new solution strategy with weight cancellation, we have selected a recent benchmark for neutron noise problems, consisting in a two-dimensional, two-group reflected fuel assembly with a single vibrating fuel pin. Approximate weight cancellation and exact regional weight cancellation were tested on this benchmark problem: use of a relatively coarse mesh with approximate cancellation leads to improvements in the FOM by factors as high as 1624 when compared to the original neutron noise scheme implemented in \tripoli{}, at the expense of a slight but undetected bias in the noise field. Use of a finer mesh for approximate cancellation, which is required in order to quench the bias inherent to this method, yielded an improvement in the FOM by factors as high as 487. By comparison, the exact weight cancellation technique only improved the FOM by a factor of 14 and showed less potential for implementation in production Monte Carlo codes, due to its algorithmic complexity.
    
    The application of approximate weight cancellation improves the quality of the resulting noise field due to mechanical vibrations to the point that it becomes feasible to use Monte Carlo methods to analyse changes in the noise field stemming from small changes to the vibration frequency and amplitude; previously, without the use of weight cancellation, such a comparison would have been extremely challenging, if not impossible, for the case of a fuel pin vibration \cite{VinaiSubmitted2022}. Our results would indicate that the application of weight cancellation to Monte Carlo neutron noise simulations is essentially mandatory if one is to obtain usable results for the study of neutron noise in power reactors. Although further investigations are required, our analysis shows that the bias imposed on the resulting noise field by the approximate weight cancellation is typically small, and more than acceptable since it can be reduced by refining the cancellation mesh. While refining the mesh will reduce the efficiency of cancellation (and also the efficiency of the noise calculation), the observed improvements in the FOM for the examined benchmark problem suggest that this is likely not a problem. In the future, we hope to demonstrate this new Monte Carlo neutron noise solution scheme on larger problems, such as vibrating fuel assemblies, or vibrating clusters of fuel pins, to further demonstrate the potential for Monte Carlo method to be applied to neutron noise analysis.
    
    \section*{Acknowledgements}

    \tripoli{} and APOLLO3$^{\text{\textregistered}}$ are registered trademarks of CEA. The authors thank EDF and Framatome for partial financial support.


\end{document}